\documentclass[preprint]{aastex7}

\usepackage{xspace}
\usepackage{graphicx}
\usepackage{verbatim}
\usepackage{listings}
\usepackage{tabularx}
\usepackage{booktabs}
\usepackage{amsmath}
\usepackage{lipsum}
\usepackage{cancel}
\usepackage{cases}
\usepackage{mathtools}
\usepackage{bm}
\usepackage{caption}
\usepackage{subcaption}
\usepackage{tikz}
\usepackage{multirow}
\usepackage{CJK}
\usepackage{hhline}

\newcommand{\update}[1]{\textcolor{black}{#1}}

\newcommand{\package}[1]{\texttt{#1}}
\newcommand{\python}{\package{Python}\xspace}

\begin{document}
\begin{CJK*}{UTF8}{bkai}

\title{Predictions of the LSST Solar System Yield: Neptune Trojans}

\author[0000-0001-9505-1131]{Joseph Murtagh}
\affiliation{Astrophysics Research Centre, School of Mathematics and Physics, Queen's University Belfast, BT7 1NN, UK}
\email[show]{jmurtagh05@qub.ac.uk}
\correspondingauthor{Joseph Murtagh}

\author[0000-0003-4365-1455]{Megan E. Schwamb}
\affiliation{Astrophysics Research Centre, School of Mathematics and Physics, Queen's University Belfast, BT7 1NN, UK}
\email{m.schwamb@qub.ac.uk}

\author[0000-0003-0743-9422]{Pedro H. Bernardinelli}
\altaffiliation{DiRAC Post-doctoral Fellow}
\affiliation{DiRAC Institute, University of Washington, 3910 15th Ave NE, Seattle, WA, USA}
\affiliation{Department of Astronomy, University of Washington, 3910 15th Ave NE, Seattle, WA, USA}
\email{phbern@uw.edu}

\author[0000-0001-7737-6784]{Hsing~Wen~Lin (林省文)}
\affiliation{Department of Physics, University of Michigan, Ann Arbor, MI 48109, USA}
\affiliation{Michigan Institute for Data and AI in Society, University of Michigan, Ann Arbor, MI 48109, USA}
\email{hsingwel@umich.edu}

\author[0009-0005-5452-0671]{Jacob A. Kurlander}
\affiliation{Department of Astronomy, University of Washington, 3910 15th Ave NE, Seattle, WA, USA}
\email{jkurla@uw.edu}

\author[0000-0001-5930-2829]{Stephanie R. Merritt}
\affiliation{Astrophysics Research Centre, School of Mathematics and Physics, Queen's University Belfast, BT7 1NN, UK}
\email{s.merritt@qub.ac.uk}

\author[0000-0002-0672-5104]{Samuel Cornwall}
\affiliation{Department of Aerospace Engineering, Grainger College of Engineering, University of Illinois at Urbana-Champaign, Urbana, IL 62802, USA}
\email{samuel41@illinois.edu}

\author[0000-0003-1996-9252]{Mario Juri\'c}
\affiliation{DiRAC Institute, University of Washington, 3910 15th Ave NE, Seattle, WA, USA}
\affiliation{Department of Astronomy, University of Washington, 3910 15th Ave NE, Seattle, WA, USA}
\email{mjuric@uw.edu}

\author[0000-0002-8418-4809]{Grigori Fedorets}
\affiliation{Finnish Centre for Astronomy with ESO, University of Turku, FI-20014 Turku, Finland}
\affiliation{Department of Physics, P.O. Box 64, 00014 University of Helsinki, Finland}
\email{grigori.fedorets@helsinki.fi}

\author[0000-0002-1139-4880]{Matthew J. Holman}
\affiliation{Center for Astrophysics $\mid$ Harvard \& Smithsonian, 60 Garden St., MS 51, Cambridge, MA 02138, USA}
\email{mholman@cfa.harvard.edu}

\author[0000-0002-1398-6302]{Siegfried Eggl}
\affiliation{Department of Aerospace Engineering, Grainger College of Engineering, University of Illinois at Urbana-Champaign, Urbana, IL 62802, USA}
\affiliation{Department of Astronomy, University of Illinois at Urbana-Champaign, Urbana, IL 61801, USA}
\email{eggl@illinois.edu}

\author[0000-0001-5916-0031]{R. Lynne Jones}
\affiliation{Rubin Observatory, 950 N. Cherry Ave., Tucson, AZ 85719, USA}
\affiliation{Aston Carter, Suite 150, 4321 Still Creek Dr., Burnaby, BC V5C6S, Canada}
\email{ljones.uw@gmail.com}

\author[0000-0003-2874-6464]{Peter Yoachim}
\affiliation{DiRAC Institute, University of Washington, 3910 15th Ave NE, Seattle, WA, USA}
\affiliation{Department of Astronomy, University of Washington, 3910 15th Ave NE, Seattle, WA, USA}
\email{yoachim@uw.edu}

\author[0000-0001-5820-3925]{Joachim Moeyens}
\affiliation{Asteroid Institute, 20 Sunnyside Ave., Suite 427, Mill Valley, CA 94941, USA}
\affiliation{DiRAC Institute and the Department of Astronomy, University of Washington, 3910 15th Ave NE, Seattle, WA 98195, USA}
\email{moeyensj@uw.edu}

\author[0009-0009-2281-7031]{Jeremy Kubica}
\affiliation{McWilliams Center for Cosmology, Department of Physics, Carnegie Mellon University, Pittsburgh, PA 15213, USA}
\affiliation{LSST Interdisciplinary Network for Collaboration and Computing Frameworks, 933 N. Cherry Avenue, Tucson AZ 85721}
\email{jkubica@andrew.cmu.edu}  

\author[0000-0001-6984-8411]{Drew Oldag}
\affiliation{LSST Interdisciplinary Network for Collaboration and Computing Frameworks, 933 N. Cherry Avenue, Tucson AZ 85721}
\affiliation{DiRAC Institute, University of Washington, 3910 15th Ave NE, Seattle, WA, USA}
\affiliation{Department of Astronomy, University of Washington, 3910 15th Ave NE, Seattle, WA, USA}
\email{awoldag@uw.edu}

\author[0009-0003-3171-3118]{Maxine West}
\affiliation{LSST Interdisciplinary Network for Collaboration and Computing Frameworks, 933 N. Cherry Avenue, Tucson AZ 85721}
\affiliation{DiRAC Institute, University of Washington, 3910 15th Ave NE, Seattle, WA, USA}
\affiliation{Department of Astronomy, University of Washington, 3910 15th Ave NE, Seattle, WA, USA}
\email{maxwest@uw.edu}

\author[0000-0001-7335-1715]{Colin Orion Chandler}
\affiliation{LSST Interdisciplinary Network for Collaboration and Computing Frameworks, 933 N. Cherry Avenue, Tucson AZ 85721}
\affiliation{DiRAC Institute, University of Washington, 3910 15th Ave NE, Seattle, WA, USA}
\affiliation{Department of Astronomy, University of Washington, 3910 15th Ave NE, Seattle, WA, USA}
\affiliation{Department of Astronomy and Planetary Science, Northern Arizona University, Flagstaff, USA}
\email{coc123@uw.edu}

\begin{abstract}

The NSF-DOE Vera C. Rubin Observatory's Legacy Survey of Space and Time (LSST), beginning full operations in late 2025, will dramatically transform solar system science by vastly expanding discoveries and providing detailed characterization opportunities across all small body populations. This includes the co-orbiting 1:1 resonant Neptune Trojans, which are thought to be dynamically hot captures from the protoplanetary disk. Using the survey simulator \texttt{Sorcha}, combined with the latest LSST cadence simulations, we present the very first predictions for the Neptune Trojan yield within the LSST. We forecast a model-dependent median number of $\sim130-300$ discovered Neptune Trojans, \update{and infer a notable 2:1 detection bias toward the recently emerged L5 cloud near the \update{galactic plane} versus the L4 cloud, reflecting the lower-cadence coverage in the Northern Ecliptic Spur region that suppresses L4 detections}. The additionally simulated Science Validation survey will offer the very first early insights into this understudied cloud. Around 60\% of detected main survey Neptune Trojans will meet stringent color light curve quality criteria, increasing the sample size more than fourfold compared to existing datasets. This enhanced sample will enable robust statistical analyses of Neptune Trojan color and size distributions, crucial for understanding their origins and relationship to the broader trans-Neptunian population. These comprehensive color measurements represent a major step forward in characterizing the Neptune Trojan population and will facilitate future targeted spectroscopic observations.

\end{abstract}


\section{Introduction} \label{sec:1}

The Neptune Trojans (NTs) are a class of minor solar system body which co-orbit with Neptune (semimajor axis $a \sim 30.1$ au), librating around the 1:1 mean-motion resonance with the giant planet. As with other Trojans, the NTs reside in the L4 (leading) and L5 (trailing) Lagrange points of Neptune. The 1:1 resonance (as with Neptune's other $n:1$ resonances) admits both symmetric (relative to the relative mean longitude $\lambda - \lambda_{N}$, \citealt{murray99, voyatzis05}; also known as a horseshoe orbit) and asymmetric (or, tadpole orbit) librators \citep{dermott81, peale86, beauge94, malhotra96, murrayclay05, gladman12, pike15, volk18, chen19}. Numerical simulations have shown that some known symmetric librators have dynamical stability on the order of $\sim$Myrs, implying they are only temporary captures from e.g. the Centaur population \citep{horner06, horner10, delafuentesmarcos12, guan12, horner12, horner12b, alexandersen13, lin16, wu19}. On the other hand, the asymmetric librators have been shown to be long-lived on the scale of $\sim$Gyrs, suggesting that these NTs are remnants of the early solar system, captured in Neptune's outwards migration \citep{marzari03, brasser04, kortenkamp04, zhou09, zhou11, lykawka11, guan12, lin16, wu19, lin21, lin22}. Recent modeling has shown that the present day NT clouds may in fact only be a small remnant of this primordial population, with only $\sim$1\% surviving after 4 Gyr \citep{lykawka09}. Most importantly, however, is that these stable NTs (both those captured and those formed in-situ, \citealt{lykawka09, lykawka11}) are therefore expected to preserve a fossil record of their source region, with their inclination $i$, eccentricity $e$, and surface color distributions remaining largely unaltered over these $\sim$Gyr timescales \citep{nesvorny09, lykawka09, lykawka10, lykawka10b, lykawka11, parker15, chen16, gomes16}. As these elements are effectively unchanged since emplacement into the resonance, the present-day orbital and color distributions can be compared quantitatively with capture and migration models of the protoplanetary disk. A well-characterized sample of NTs could therefore assist in discriminating between competing migration scenarios \cite[e.g.,][]{nesvorny12} and inform the early dynamical evolution of the outer solar system.

The known NT population is, however, relatively small, being composed from discoveries by several surveys \citep{elliot05, sheppard06, sheppard10, parker13, alexandersen13, gerdes16, bannister18, lin19, bernardinelli22}. The Minor Planet Center (MPC) only lists 31 NTs at the time of writing\footnote{\href{https://www.minorplanetcenter.net/iau/lists/NeptuneTrojans.html}{https://www.minorplanetcenter.net/iau/lists/NeptuneTrojans.html}}, with the vast majority being located within the L4 cloud (27 L4 vs 4 L5). This asymmetry is almost certainly due to the observational bias of the L5 cloud having resided in the crowded \update{galactic plane} for at least the past two decades, making the discovery of L5 NTs extremely challenging. The first discovery of an L5 NT by \cite{sheppard10} implied a symmetrical distribution of L4 and L5 clouds, with numerical simulations from \cite{chiang05} predicting them to be at least as numerous as the Jupiter Trojans. Many of the known L5 NTs have been shown to be dynamically unstable (2013 KY$_{18}$, \citealt{lin16}; 2004 KV$_{18}$, \citealt{guan12, horner12b}; 2008 LC$_{18}$, \citealt{horner12}), therefore prompting that these are temporary captures, reinforcing the potential for asymmetric cloud sizes. Although relatively few NTs have been detected to date, limiting the precision of population-wide constraints, ongoing studies have begun to characterize their total population sizes, size-frequency distribution, and surface colors. Early color measurements initially suggested that NTs may share similarly bimodal surface color distributions as with the Jupiter Trojans \citep{sheppard06, sheppard12, parker13, gerdes16, jewitt18, lin19, bolin23}, although more recent surveys from \cite{markwardt23} and \cite{bolin23} show that this may be a result of a lack of observations. Long-term collisional and dynamical models show that impacts between NTs and the e.g. Plutino population, whilst relatively rare over Gyr timescales \citep{almeida09}, may contribute to additional fragmentation and shared surface colors of both populations \citep{thebault03, deelia08, almeida09, nesvorny09, lin21, markwardt25}. Whilst NTs are clearly a potentially highly informative population, current observations provide only a small, biased glimpse of their true underlying distribution.

The NSF-DOE Vera C. Rubin Observatory's Legacy Survey of Space and Time \citep[LSST;][]{lsst09, ivezic19, bianco22} will dramatically advance our knowledge of the entire solar system. With its effective 6.5m aperture and \update{$\sim$9.6 deg$^2$} field of view, it will image the entire southern night sky approximately every three nights in six optical $ugrizy$ bands - all over a 10-year observational baseline. The `wide-fast-deep' (WFD) portion of the survey will cover \update{$\sim$18,000 deg$^2$} repeatedly (with $\sim800$ visits per field) to a single visit depth of $m_r \sim 24.7$ (or $\sim27.5$ in co-added images; \citealt{lsst09, ivezic19, bianco22}). Per \cite{jones20} and \cite{scocv3}, the LSST will also allocate $\sim$10\% of its observing cadence time to mini-surveys. In particular for solar system science, the largest components include the Deep Drilling Fields (DDF; 6 dedicated \update{$\sim$10 deg$^2$} survey areas which will receive a higher cadence of visits to achieve deeper co-add magnitude depths) and the Northern Ecliptic Spur (NES; an extension in $griz$ observations of the WFD by $+10^\circ$ ecliptic latitude; \citealt{schwamb18, schwamb23}). The combination of its wide field of view, faint sensitivity, multi-band imaging, and dense observational cadence make the LSST particularly suited to detecting solar system objects. Prior studies have predicted that the near-Earth objects, \update{main-belt} asteroids, Jupiter Trojans, and trans-Neptunian objects (TNOs) will all receive a $\sim$10 fold increase in their respective populations \citep{jones09, lsst09, solontoi10, shannon15, grav16, silsbee16, veres17, jones18, eggl19, ivezic19, fedorets20, hoover22}. More recently, \cite{kurlander25} more precisely predict $\sim4-10$ fold for these populations, with \cite{murtagh25} predicting a $\sim$7 fold increase for the Centaurs. It is logical to expect that the LSST will therefore also be capable of vastly increasing the known NT sample size as it not only probes down to depths of \update{$m_r \sim 24.7$}, but also tiles more of the entire southern night sky across multiple filters than previous surveys that have discovered NTs \citep[e.g.,][]{chiang03, sheppard06, becker08, sheppard10b, gerdes16, lin16, bannister18, lin19, bernardinelli20, bernardinelli22}.

In this work, we present the very first estimates of the NT population yield within the LSST, quantifying the expected discovery numbers, characterization potential through phase and light curves, and implications for study of the NT population. Forecasting the discoveries of NTs will be vitally important in the LSST era for aiding in planning follow-up observational campaigns for e.g., orbit refinement or spectroscopy, ensuring that the LSST is fully exploited. Such predictions will also allow for tests to be performed on NT population models - different capture and migration scenarios produce differing NT distributions, so comparison with real LSST samples will help constrain these scenarios. Knowledge of the NT inclination distribution shape may help in distinguishing between capture scenarios from a widely dispersed disk with cold/hot components or a single component thinner width disk \citep{nesvorny09, lykawka09, parker15, chen16, lin21}. Additionally, numerical simulations show that 3\% of Centaurs may originate from formerly captured NTs, producing an additional reservoir that can be understood by measurement of the NT population \citep{horner10}. Given the current lack of constraints on the NT population, the LSST's immense observational capabilities, and its imminent late-2025 commencement, it is therefore timely to estimate just how many NTs the LSST should be able to detect. 

In Section \ref{sec:2}, we describe the setup for our simulation, including a description of the LSST cadence simulation, the survey simulator \texttt{Sorcha} \citep{merritt25, holman25}, and our models for the NTs. In Section \ref{sec:3}, we outline the major results for the NT discoveries within the decade-long operation of the LSST, including the expected timelines of discoveries and number of observations of both L4 and L5 clouds, and the light curve and phase curve opportunities. Section \ref{sec:4} looks at the predictions for discovery yields within the three-month-long science validation survey. Finally, Section \ref{sec:5} summaries these results and discusses factors that may influence the interpretation and scope of these findings.

\section{Method} \label{sec:2}

In order to estimate the NT population observable within the LSST, we set out to build a model for the NTs, as has similarly been done previously in \cite{kurlander25} and \cite{murtagh25}, using the orbital models described in \cite{alexandersen16} and \cite{lin16, lin21}. We use the survey simulator \texttt{Sorcha} \citep{merritt25, holman25}, combined with the most up-to-date simulation of the LSST's pointing history and model observing conditions in order to create a biased set of observations. From these, we investigate the total number of objects that are detectable within the LSST, when these objects are detectable, and where on-sky these detections occur. We look to what datasets the LSST will be able to provide for phase curve analysis, surface colour measurements, and cloud symmetry.  

\subsection{Sorcha} \label{sec:2.1}
Our NT observations are simulated using the open-source, modular survey simulator \texttt{Sorcha} \citep{merritt25, holman25}. We refer the interested reader to \cite{merritt25} for a full discussion on the functionality of \texttt{Sorcha}, but provide a brief description here for completeness. \texttt{Sorcha} ingests an input model for the NTs, based on their orbital elements, absolute magnitudes in $r$ band and optical surface colors with respect to $r$ (as well as optional phase curve parameters). It then generates ephemerides for all objects with its \update{built-in} $n$-body integrator ASSIST \citep{holman23}, itself an extension of REBOUND \citep{rein12, rein15} - see \cite{merritt25} for a list of massive perturbers used. Given a description of the Rubin Observatory (including its position, camera footprint, and realistic weather conditions), and the LSST survey itself (the filters used, the exposure times, where and when on-sky it will observe), \texttt{Sorcha} then evaluates the detectability of objects in each exposure. Detectability is determined from distance to the center of the pointing center, the exact footprint of the LSSTCam (including gaps in the CCD), removal of saturated objects, and an application of a detection efficiency function as per \cite{chelsey17}. Note that initially we assume a uniform detection efficiency across the entire sky - we discuss the implications of crowded stellar fields such as the \update{galactic plane} on NT detection in Section \ref{sec:3.3}. After assessing object detectability, \texttt{Sorcha} then emulates the Rubin Solar System Processing (SSP) object linking algorithm, which is designed to associate three nightly tracklets - separated by more than 5$^{\prime\prime}$ spatially and less than 90 minutes temporally - within a 15-day window, all with an expected 95\% linking success rate \citep{lsst09, ivezic13, myers13, ivezic19, juric20}. The final database of detections produced by \texttt{Sorcha} can therefore be used to provide a realistic assessment of the numbers of discovered and linked objects, as well as any characterization metrics such as light curves, phase curves, or surface colors.

\subsection{The LSST Cadence Simulations\label{sec:2.2}}
\begin{figure*}
    \centering
    \includegraphics[width=\textwidth]{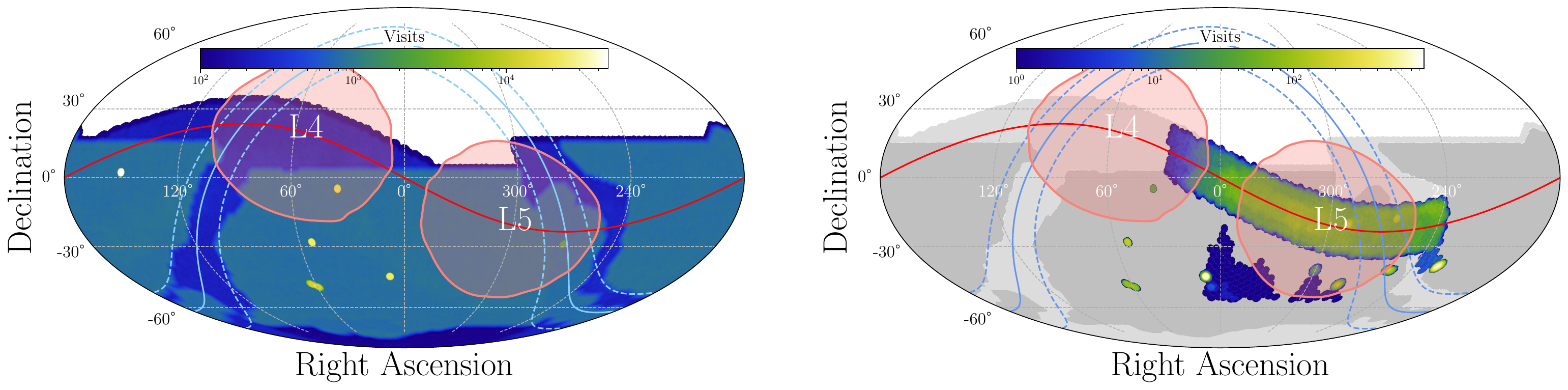}
    \caption{Skymaps of the number of visits across all filters of (\textit{left}) the 10 year LSST survey cadence based on the one snap v4.3.1 simulation \citep{scocv3} and (\textit{right}) the Science Validation survey based on the \update{\texttt{lsstcam\_20250930}} simulation \citep{claver25, sv0903} — note the difference in color scale between the two. The main wide-fast-deep survey makes up $\sim$80\% of the total main LSST survey time, receiving on average $\sim$800 visits per pointing. Also of note are the Northern Ecliptic Spur (NES) \update{mini-survey} (\update{dark blue region}, $+10^\circ$ ecliptic latitude) and the Deep Drilling Fields (yellow-white circles, regions with high temporal sampling cadence). For full details of the survey strategy, see \cite{scocv3}. Overplotted on both in solid red is the ecliptic plane. In solid blue is the \update{galactic plane}, with the corresponding dashed cyan lines representing $\pm10^\circ$. \update{On top of all of this in pink are the positions of both L4 and L5 clouds at the beginning of the survey for context.}}
    \label{fig:cadence}
\end{figure*}

Our model simulation uses the ``one\_snap" cadence from v4.3.1 of the observing strategy detailed in the Survey Cadence Optimization Committee (SCOC) report \cite{scocv3} and with a skymap of the distribution of visits in all filters shown in Figure \ref{fig:cadence}. This simulation has been generated by the \texttt{rubin\_sim} \citep{bianco22, yoachim23} and \texttt{rubin\_scheduler} \citep{naghib19, yoachim24b} packages\footnote{For the most up-to-date cadence simulations, see \hyperlink{https://s3df.slac.stanford.edu/data/rubin/sim-data/}{https://s3df.slac.stanford.edu/data/rubin/sim-data/}}. It accounts for the assumed weather conditions, optics system response (including telescope and camera performance, and mirror coating specifications), and individual filter responses \citep{connolly14, delgado14, delgago16, yoachim16, lsst17, jones18, jones20, naghib19, bianco22}. Compared to prior realizations of the cadence which employed 2x15s exposures, observations are obtained in a single 29.2s exposure (or 38s in $u$) as commissioning tests of the LSSTCam have shown that cosmic ray removal is feasible in single visit observations. Compared to prior iterations of the cadence simulation, v4.3.1 includes an update to when the LSST will start observing in November 2025 \citep{guy21}. This will prove to be highly important in analyzing the detection timeline and discovery rates of the highly localized on-sky positions of the NT population. For comparison to \cite{murtagh25}, we ran 100 simulations using the v4.0 cadence simulation \citep{scocv3}, which has a prior assumed start date of May 2025, as well as a redistribution in visits throughout year 1 to enable template generation. This resulted in a negligible difference in the final reported yield, only slightly affecting the timing of discoveries. 

In addition to simulating the main LSST cadence, we also simulate the Science Validation (SV) survey \citep{claver25} as also shown in Figure \ref{fig:cadence}. The SV survey is an early phase of LSST operations, commenced in June 2025 for three months of observing, designed to validate and refine system performance (e.g., alert generation, data release processing pipelines, solar system discovery pipelines) before full science operations commence. The overall image quality of the SV is expected to be worse than the main survey due to degraded seeing and weather conditions, with a smaller, focused \update{$\sim750$ deg$^2$} area than initially planned as a result. \update{The simulated cadence file we use, \texttt{lsstcam\_20250930} \citep{sv0903}, utilizes the actual on-sky observations taken since commissioning commenced, with any non-science grade observations removed.} Simulating the SV survey allows us to assess early-year discovery probabilities under its limited and uneven footprint and understand how early coverage and cadence choices will shape the initial characterization of the NT population. 

\subsection{Neptune Trojan Model\label{sec:2.3}}
Our NT model is split into three components; an orbital model consisting of semi-major axis $a$, eccentricity $e$, inclination $i$, libration amplitude $L_{1:1}$, argument of perihelion $\omega$, longitude of the ascending node $\Omega$, and mean anomaly $M$; and a surface colors model, and a model of absolute magnitudes $H$ with respect to $r$ band. We do not include any treatment for phase curves, as \cite{kurlander25} has shown that the distant TNO movement of $\sim0^\circ-3^\circ$ in phase angle is approximately linear, and leads to minimal effects on object brightness and so observability within the LSST \citep[\update{cf.} phase curve measurements of TNOs from][]{dobson23}. In addition, previous NT surveys focused primarily on single color measurements rather than phase curves of NTs \citep[e.g.,][]{sheppard06, sheppard10, lin19, bolin23, markwardt23}, and so any phase curve parameter is poorly constrained. The input parameters for each model in this work are summarized in Table \ref{tab:allparams}. We detail our absolute magnitude, color, and orbital models in the following subsections.

\begin{deluxetable}{c | c c c c c c}
\tablecaption{Input NT population model parameters for the three models investigated in this work. \label{tab:allparams}}
\tablewidth{\textwidth}
\tablehead{
  & Parameter & Distribution type & Single-component & \multicolumn{2}{c}{Two-component} & Rolling power law \\ \cline{5-6}
  &         &                   &                  & Cold & Hot &
}
\startdata
    \multirow{8}{*}{Orbits}     & $a$ (au)                  & Uniform                    & 30.0 - 30.2 & 30.0 - 30.2 & 30.0 - 30.2 & 30.0 - 30.2 \\
                                & $e$                       & Rayleigh                   & $\sigma_e = 0.045$ & $\sigma_e = 0.02$ & $\sigma_e = 0.05$ & $\sigma_e = 0.045$ \\
                                & $i$ ($^\circ$)            & sin($i$) $\times$ Gaussian\tablenotemark{a} & $\sigma_i=14$ & $\sigma_i=6$ & $\sigma_i=18$ & $\sigma_i=14$ \\
                                & $\omega$ ($^\circ$)       & $\omega = \phi_{1:1}  - M - \Omega + \lambda_N$                    & 0 - 360 & 0 - 360 & 0 - 360 & 0 - 360 \\
                                & $\Omega$ ($^\circ$)       & Uniform                    & 0 - 360 & 0 - 360 & 0 - 360 & 0 - 360 \\
                                & $M$ ($^\circ$)            & Uniform                    & 0 - 360 & 0 - 360 & 0 - 360 & 0 - 360 \\
                                & $\phi_{1:1}$ ($^\circ$)   & Equation \ref{eq:11}       & 0 - 360 & 0 - 360 & 0 - 360 & 0 - 360 \\
                                & $L_{1:1}$ ($^\circ$)      & Rayleigh                   & $\sigma_{L_{1:1}} = 12$ & $\sigma_{L_{1:1}} = 12$ & $\sigma_{L_{1:1}} = 12$ & $\sigma_{L_{1:1}} = 12$ \\
    \hline
    \multirow{2}{*}{Absolute magnitude} & \multirow{2}{*}{$H_r$} & \multirow{2}{*}{Power law} & Divot & Single slope & Divot & Rolling \\ 
    & & & (0.9, 0.4, 8.3, 3.2)\tablenotemark{b} & (0.2, 8)\tablenotemark{c} & (0.9, 0.4, 8.3, 3.2)\tablenotemark{b} & (0.844, -0.227)\tablenotemark{d}
    \\
    \hhline{=======}
    & \multirow{2}{*}{Color component} & \multicolumn{2}{c}{\multirow{2}{*}{Mean $\pmb{\mu}$}} & \multicolumn{2}{c}{\multirow{2}{*}{Covariance \pmb{C}}} & \multirow{2}{*}{Mixing weight $f_\alpha$} \\
     & & & & & \\  
    \cline{2-7}
    \multirow{2}{*}{Color\tablenotemark{e}} & NIRB\tablenotemark{e} & \multicolumn{2}{c}{$\begin{pmatrix} 0.609 & 0.223 & 0.355 \end{pmatrix}$} & \multicolumn{2}{c}{$\begin{pmatrix} 0.00434 & 0.00337 & 0.00547 \\ 0.00337 & 0.00268 & 0.00437 \\ 0.00547 & 0.00437 & 0.00727 \end{pmatrix}$} & 0.682 \\
      & NIRF\tablenotemark{e} & \multicolumn{2}{c}{$\begin{pmatrix} 0.902 & 0.347 & 0.524 \end{pmatrix}$} & \multicolumn{2}{c}{$\begin{pmatrix} 0.00674 & 0.00362 & 0.00557 \\ 0.00362 & 0.00205 & 0.00321 \\ 0.00557 & 0.00321 & 0.00533 \end{pmatrix}$} & 0.318 \\
 \enddata
 \tablecomments{\update{$\lambda_N$ is Neptune's mean longitude. $\sigma_{e/i/L_{1:1}}$ are the widths of the associated distribution. They are discussed in further detail in Section \ref{sec:2.3.3}.}}
 \tablenotetext{a}{Truncated at $60^\circ$}
 \tablenotetext{b}{($\alpha_b$, $\alpha_f$, $H_B$, $c$); the bright end slope parameter, the faint end slope parameter, and the break magnitude, and the contrast parameter respectively}
 \tablenotetext{c}{($\alpha_1$, $H_{max}$); the cold component slope parameter, and the cutoff magnitude respectively}
 \tablenotetext{d}{($\theta$, $\theta$'); the slope curvature parameter, and first derivative of the slope curvature parameter respectively}
 \tablenotetext{e}{All three absolute magnitude models share the same GMM-based color model from \cite{bernardinelli25}. Here, NIRB is the near-infrared bright component and NIRF is the near-infrared faint component (see Section \ref{sec:2.3.2})}
\end{deluxetable}   
    
\subsubsection{Absolute Magnitude Distribution} \label{sec:2.3.1}
\begin{figure}
    \centering
    \includegraphics[width=\linewidth]{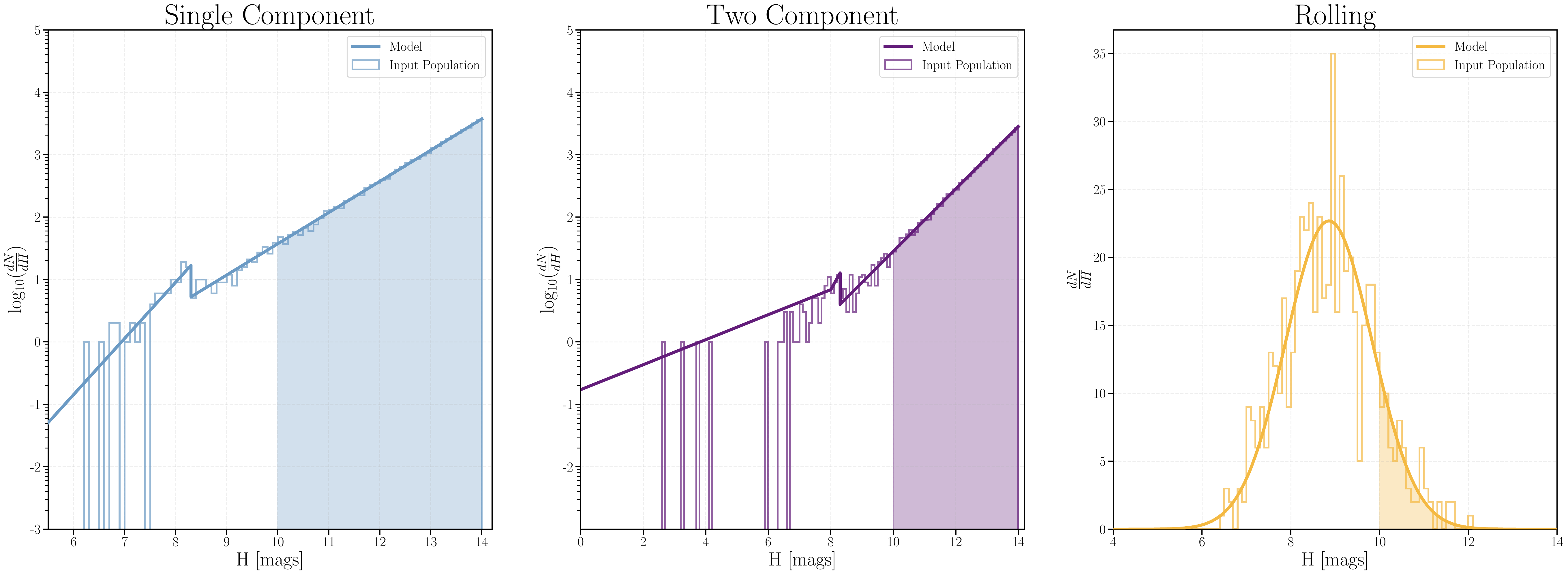}
    \caption{Differential histograms of all 3 absolute magnitude $H$ distributions used in modeling the NT population in this work. In all cases, the solid line represents the analytical model for the distribution, whilst the histograms are the uniformly sampled $N_{NT}$ objects from that model. \update{The shaded regions in all cases represent the region beyond the LSST's nominal single-visit detection threshold ($H_r \approx 10$) for an object at Neptune's orbit.} Represented here are \textit{(left panel)} the single-component model, \textit{(middle)} the two-component model, and \textit{(right)} the rolling power law \update{model}. \update{Note that the single- and two-component models have a logarithmic y-axis scale, whilst the rolling power law model is linear.}}
    \label{fig:allH}
\end{figure}

Determining a size distribution requires knowledge of object geometric albedo; yet for NTs, albedos are both sparsely measured \citep{markwardt23} and may be correlated with surface color as in other TNO populations \citep[e.g.,][]{lacerda14b, fraser14}. Rather than introducing poorly constrained systematics, we model the absolute magnitude $H$ distribution as a proxy for the convolution of size and albedo, which is directly measurable from survey photometry. Motivated by a growing consensus in related TNO studies, we evaluate three candidate $H$ distributions \update{(highlighted in Figure \ref{fig:allH})}:

\begin{enumerate}
    \item A single-component `divot' model \citep{lin21} that introduces a discontinuity at a break magnitude $H_B$ quantified by a contrast parameter $c$ before another exponential rise. This is indicative of a population that has been `frozen-in' after being scattered from an earlier collisional environment \citep{fraser09, shankman13, shankman16, alexandersen16}.
    \item A two-component `divot' model \citep{lin21}, whereby there is a size-dependent `cold' and `hot' component. This distribution may reflect either a primordial two-component population captured during Neptune's migration \citep{lykawka09, lykawka11}, a single origin later split by Plutino collisions, or the emergence of a cold collisional family.
    \item A rolling power law \citep{bernstein04, bernardinelli25}, whose logarithmic slope evolves smoothly via a curvature parameter $\theta$. \cite{bernardinelli25} show that, for the Dark Energy Survey (DES) sample of 696 TNOs, this model can provide a common fit for all TNO dynamical and color families over the measured size range, in line with results from \cite{petit23}. This therefore suggests a common origin population, similar to the color families from \cite{fraser23}.
\end{enumerate}

With only 31 $H$ measurements recorded in the MPC, as well as previously mentioned albedo-color correlation ambiguities, the intrinsic $H$ distribution of NTs remains effectively unconstrained, and so may be described by any of the above models. For the NTs (and other scattering objects), only a single slope $H$ distribution has been previously disfavored by \cite{sheppard10b, shankman13, lin21} due to a deficit of intermediate sized NTs ($H \approx 9-11$). As a result, we investigate the above three distributions in our LSST predictions in order to probe a more complete range of NT size-distribution behavior. The input parameters for each $H_r$ model are highlighted in Table \ref{tab:allparams}.

To set our maximum simulated $H_r$ value in all cases, we take the faintest recorded $griz$ visit magnitudes within the v4.3.1 cadence simulation, and calculate the value of $H$ at this apparent magnitude at 30.1 au. Each $H$ value is converted to a $H_r$ value using the brightest $g-r/i-r/z-r$ colors obtained from the distributions in Section \ref{sec:2.3.2}. Finally, an additional margin of two magnitudes is applied in order to account for the fact that the limiting magnitude represents 50\% detection efficiency, and so repeat observations of faint objects slightly fainter than the limiting magnitude may be observable with enough repeat observations. This results in a faint end cutoff of $H_{max} = 14$, for which we generate down to for all distributions described following.

\paragraph{Single-Component Model}
The single-component model as detailed in \cite{lin21} describes two single power laws separated by a `divot' component at some break magnitude $H_B$, with differing slopes on the bright and faint ends of the distribution. The cumulative total number of NTs for a faintest absolute magnitude $H_{max}$ after $H_B$ is given by the following equation (see Appendix \ref{ap:1} for full derivation):

\begin{equation} \label{eq:1}
    N({\leq}H_{max}) = \underbrace{k_b 10^{\alpha_b H_B}}_{H < H_B} + \underbrace{k_b \frac{1}{c} \frac{\alpha_b}{\alpha_f} \frac{10^{\alpha_b H_B}}{10^{\alpha_f H_B}} \left( 10^{\alpha_f H_{max}} - 10^{\alpha_f H_B}\right)}_{H \geq H_B}
\end{equation}
where $k_b$ is a normalization constant that controls the size of the distribution, $\alpha_b/\alpha_f$ are slope values for the bright and faint ends of the distribution respectively, and $c$ is a contrast parameter that controls the size of the post-divot drop. By performing an inverse transform of Equation \ref{eq:1}, the number of objects to be drawn from this total distribution can therefore be assigned $H$ values using the following equations, according to the ratio of bright end to total objects (again see Appendix \ref{ap:1} for a full derivation and determination of this ratio):

\begin{equation} \label{eq:4}
    H = 
    \begin{cases}
        \frac{1}{\alpha_b} \left( \log_{10}(\mathcal{U}) + \alpha_b H_B \right) & H < H_B \\
        \frac{1}{\alpha_f} \log_{10} \left( \mathcal{U} \cdot \left( 10^{\alpha_f H_{max}} - 10^{\alpha_f H_B} \right) + 10^{\alpha_f H_B} \right) & H \geq H_B
    \end{cases}
\end{equation}
where $\mathcal{U}$ is a random number in the uniform distribution $\in [0, 1)$. We use the values for ($\alpha_b, \alpha_f, c, H_B$) = (0.9, 0.5, 3.2, 8.3) from \cite{lin21}, as determined in \cite{lawler18} before, as they have already proven to be non-rejectable best fits to OSSOS NT data using Anderson-Darling sampling tests. We use a population scaling from \cite{lin21}, who use the Pan-STARRS1 and OSSOS+ detections with a survey simulator to estimate (assuming symmetric L4 and L5 clouds) that there are $360^{+230}_{-196}$ NTs with $H < 10$ (in agreement with previous estimates from \citealt{lin19}). From this we find that this model produces $N_{NT}(H<14) = 32,472$. This model $H$ distribution also produces $N_{NT}(H<7)=5$ objects. Whilst the MPC contains only one such object with $H<7$, the apparent magnitude of such an object would approximately be $m\sim21.7$, which is well within the LSST's nominal limiting depth of $m_r \sim 24.7$ \citep{bianco22}. As the survey space of NTs has been relatively limited thus far \update{and remains incomplete at the bright end as shown from prior Pan-STARRS1 and OSSOS+ surveys \citep[][]{sheppard10b, lin16, lin21}}, combined with much larger objects existing within other resonances (e.g., Pluto with $H_V=-0.53$ in the 3:2 resonance, as per the MPC), we do not completely rule out the possibility of such future discoveries. However, we note that \citet{bernstein04, brown08, petit11, kavelaars21, napier22} all report a potential lower size limit of $H=5.65$ for dynamically cold objects. If applicable to NTs, this would imply that our model's prediction of zero such objects is consistent with these dynamical constraints. The resulting $H$ distribution, after sampling $N_{NT}$ many objects using Equations \ref{eq:4}, is shown in Figure \ref{fig:allH}.

\paragraph{Two-Component Model}
\cite{lin21} also find significant statistical evidence for a two-component $H$ distribution of the NTs, whereby the bimodal $H$ distribution can be described with a single power law `cold' component, and a divot power law `hot' component as follows:

\begin{equation}
    N(\leq H_{max}) =
        \left\{
            \begin{array}{@{}l@{\quad}l}
                \text{Cold:} &
                    \begin{array}{l}
                        \underbrace{k_1 10^{\alpha_1 H_{max}}}_{H < 8} 
                    \end{array}
                 \\[1.5ex]
                \text{Hot:} &
                    \begin{array}{l}
                       \underbrace{k_b 10^{\alpha_b H_B}}_{H < H_B} + \underbrace{k_b \frac{1}{c} \frac{a_b}{a_f} \frac{10^{a_b H_B}}{10^{af H_B}} \left( 10^{a_f H_{max{}}} - 10^{a_f H_B} \right)}_{H \geq H_B} 
                    \end{array}
            \end{array}
        \right.
\end{equation}
where $k_1 / k_b $ are normalization constants which control the size of the distribution for the cold and hot component respectively, $\alpha_1 / \alpha_b / \alpha_f$ are the slope values for the cold $H<8$, and hot $H < H_B$, and $H \geq H_B$ parts of the distribution respectively, and $c$ is a contrast parameter representing the size of the drop post-divot. The number of objects to be drawn from this distribution can again be assigned $H$ values using the same inverse transform method as for Equation \ref{eq:1}:

\begin{equation}
    H =
        \left\{
            \begin{array}{@{}l@{\quad}l}
                \text{Cold:} &
                \left\{
                    \begin{array}{l}
                        \frac{1}{a_1} \left( \log_{10}(\mathcal{U}) + \alpha_1 8 \right) \quad H < 8 \vphantom{k_f \alpha_f \ln(10)\, 10^{\alpha_f H}}
                    \end{array}
                \right. \\[1.5ex]
                \text{Hot:} &
                \left\{
                    \begin{array}{l}
                        \frac{1}{\alpha_b} \left( \log_{10}(\mathcal{U}) + \alpha_b H_B \right) \quad H < H_B \\
                        \frac{1}{\alpha_f} \log_{10} \left( \mathcal{U} \cdot \left( 10^{\alpha_f H_{max}} - 10^{\alpha_f H_B} \right) + 10^{\alpha_f H_B} \right) \quad H \geq H_B
                    \end{array}
                \right.
            \end{array}
        \right.
\end{equation}
where again $\mathcal{U}$ is a random number in the uniform distribution $\in[0,1)$.

For the cold component of this population, we use the values of ($\alpha_1$, $\alpha_b$) = (0.2,0.9) as per estimates from \cite{lin21}, who find there should be $N_{cold} = N(H<8) = 26^{+22}_{-12}$ cold NTs. For the hot component of this distribution, we again use the values of ($\alpha_b, \alpha_f, c, H_B$) = (0.9, 0.5, 3.2, 8.3) from \cite{lin21} and \cite{lawler18} before, giving an estimate of $N(H<10) = 272^{+168}_{-150}$. The total number of hot component NTs to simulate is thus $N_{hot} = N(H<14) = 24,534$. This model therefore overall produces $N_{NT}(H<14) = N_{cold} + N_{hot} = 24,560$ NTs. Similarly to the single-component model, the parameters used here produce $N_{cold}(H<7)=16$ and $H_{hot}(H<7)=4$ objects. We also note that the cold components yields 11 objects with the lower-size limit of $H<5.65$ — this is however a small fraction of the discoverable population that are isolated in Section \ref{sec:3.1}, and it remains unclear whether this \update{constraint}, derived from dynamically cold TNO populations, applies directly to NTs as well. The hot and cold $H$ distributions are shown in Figure \ref{fig:allH}.

\paragraph{Rolling Power Law Model}
\update{From \cite{bernardinelli25}, the rolling power law formulation describes the probability for a given absolute magnitude $H$ within two compositional surface types (excluding the Haumea collisional family), $\beta \in \{$NIRB,NIRF$\}$, where NIRB (`near-infrared bright') and NIRF (`near-infrared faint') refer to objects that are respectively brighter or fainter in the near-infrared at similar optical colors (see Section \ref{sec:2.3.2} for further discussion). The probability distribution of the rolling power law model is given by:}

\begin{equation} \label{eq:8}
    p_{\beta}(H|\theta_{\beta},\theta_{\beta}') \propto 10^{\theta_{\beta}(H-7) + \theta_{\beta}'(H-7)^2}
\end{equation}
where $\theta_{\beta}$ and $\theta_{\beta}'$ are slope curvature parameters. As with analysis from \cite{bernardinelli25}, we assume that the NT population share the same form of rolling power law as all other dynamical families in the TNO population, which is reasonable given their observed inclination and eccentricity distributions. As such, we use the more simple (but still statistically strong, $\mathcal{R}=2.370$) assumption that the posterior distributions of the NIRB/NIRF $\theta_{\beta}$ and $\theta_{\beta}'$ overlap enough to be common, with $\theta_{NIRB} = \theta_{NIRF} = 0.844$ and $\theta_{NIRB}' = \theta_{NIRF}' = -0.227$. \cite{bernardinelli25} restricted their analysis to their detected DES sample range of $5.5 < H < 8.2$, however the MPC reports the majority (18) of known NTs having a measured $H > 8.2$. As such, we choose to extend the distribution range of Equation \ref{eq:8} by first normalizing by the debiased estimated number of NTs in this sample, $N_{NT}=139^{+78}_{-56}$, and the maximum posterior estimate in the original $H$ range as follows:

\begin{equation}
    N(<H_{max}) = \frac{N_{NT}}{\int_{5.5}^{8.2} \sum_{\beta} f_{\beta} p_{\beta}(H|\theta_{\beta},\theta_{\beta}') dH} \cdot \int_{0}^{H_{max}} \sum_{\beta} f_{\beta} p_{\beta}(H|\theta_{\beta},\theta_{\beta}') dH
\end{equation}
where $f_{\beta}$ are the fractions of the NT population who belong to the NIRB or NIRF populations (see Section \ref{sec:2.3.1}). Using this formulation, we find that this model produces $N_{NT}(H<14) = 556$ NTs with $N(H<7) = 15$\update{, and $N(H<10) = 488$)}. We uniformly sample by interpolating the cumulative distribution function of this distribution with a cubic spline fit via \texttt{scipy.interpolate.interp1d} function, resulting in the drawn $H$ distribution shown in Figure \ref{fig:allH}. The large discrepancy in $N_{NT}$ compared to the single and two component $H$ models is due to the power law index rolling over to 0 at $H\approx9$ (\update{cf.} performing a Taylor expansion of the power term in Equation \ref{eq:8}), resulting in the slope \textit{decreasing} after this. Whilst this may potentially have the caveat of underestimating the faint end of the NT distribution, there are not yet enough measurements to confidently constrain this distribution. With the sample size available from \cite{sheppard10b}, \cite{shankman12} note that the NT size distribution is consistent within 2$\sigma$ with negative slopes for $H > 9$, in line with the turnover of this rolling power law distribution. Furthermore, at the LSST's limiting magnitude of $m_r\sim24.7$, an object at 30.1 au would correspond to $H_r\approx10$, indicating that the range considered here is adequately probing the magnitude range in which the LSST is expected to discover NTs.

\subsubsection{Colors} \label{sec:2.3.2}
\begin{figure}
    \centering
    \includegraphics[width=\columnwidth]{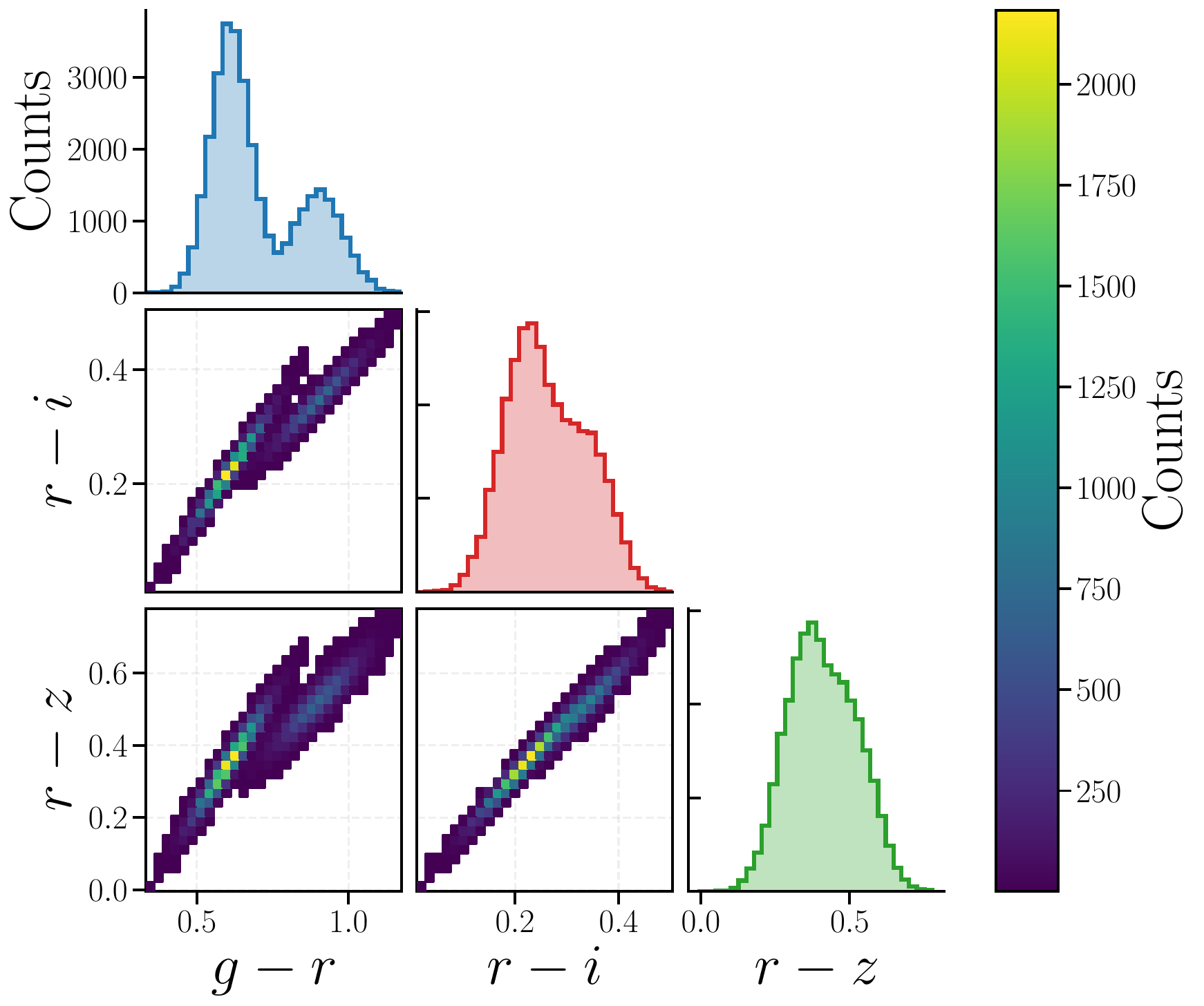}
    \caption{$g-r$, $r-i$, and $r-z$ color histograms for the input model produced from GMM parameters in Table \ref{tab:allparams}. The leading diagonal represents the individual 1D histograms of each color, whilst the middle left, bottom left, and bottom middle 2D histograms highlight their correlation. Samples are drawn to ensure this correlation is kept.}
    \label{fig:gmmcolors}
\end{figure}

TNOs, including NTs, exhibit a pronounced color bimodality, commonly partitioned into ``red'' and ``very red'' groups, which likely reflect compositional gradients in the primordial protoplanetary disk \citep{fraser12, dalleore13, schwamb19, buchanan22, fraser23, marsset23, pike23}. In particular, the optical and near-infrared colors of NTs are of interest due to their dynamically stable and collisionally unevolved status \citep{bottke23}, making them potentially direct tracers of the disk's original chemical stratification. Recently, \cite{fraser23} formalized these color distinctions into near-infrared bright (BrightIR) and near-infrared faint (FaintIR) classes, defined by their relative brightnesses in optical and near-infrared $J$-band colors, again thought to correspond to distinct primordial surface materials. Building on this framework, \cite{bernardinelli25} modeled the joint color distribution of the entire Dark Energy Survey (DES) sample (696 TNOs, including 6 NTs) as a two-component Gaussian mixture model (GMM; we refer the reader to \citealt{bernardinelli25} for a full treatment of their GMM modeling process). These components, which they refer to as \update{NIRF} and \update{NIRB}, defined by their relative $i/z$-band brightnesses at fixed $g-r$ colors, are seen to closely mirror the FaintIR and BrightIR populations in \cite{fraser23} respectively, as well as being consistent with recent James Webb Space Telescope spectral groupings of TNOs and NTs \citep{markwardt25}. We adopt their GMM parameterization to describe our NT color distribution — for a measurement $\bm{x} = (g-r, r-i, r-z)$, the color distribution it is drawn from is modeled as a weighted sum of multivariate Gaussians $\mathcal{N}$:

\begin{equation}
    p(\bm{x}|\bm{\mu}_\alpha, \bm{C}_\alpha) = \sum_{\alpha=1}^{K} f_\alpha \mathcal{N}(\bm{x}|\bm{\mu}_\alpha, \bm{C}_\alpha) \hspace{1.5em} \left(\text{where } \mathcal{N}(\bm{x} | \bm{\mu_\alpha}, \bm{C_\alpha}) \equiv \frac{1}{\sqrt{(2\pi)^3 |\bm{C_\alpha}|}} e^{-\frac{1}{2} (\bm{x} - \bm{\mu_\alpha})^\text{T} \bm{C_\alpha^{-1}} (\bm{x} - \bm{\mu_\alpha}) }\right)
\end{equation}
where $\bm{\mu_\alpha}$/$\bm{C_\alpha}$ are the mean and covariance of the $\alpha^{th}$ Gaussian of the model, weighted by $f_\alpha$ (such that $\sum_\alpha f_\alpha = 1$). From this, we utilize their estimate of the mean and covariances of the best-fit GMM output (highlighted in Table \ref{tab:allparams}), and fraction of NTs in the NIRF class (obtained from a sample size of 6 NTs) of $f_{NIRB} = 0.682 \pm 0.156$ for our color model. We then select $N_{NT}$ samples uniformly from this GMM distribution (where $N_{NT}$ is the number of NTs in a given model population — see Appendix \ref{ap:2} for an explanation of the implementation of multivariate normal sampling), ensuring that correlation in color is maintained in each selection. The resulting model color distributions are shown in Figure \ref{fig:gmmcolors}. 

We note that we do not model $u/y$-band colors, motivated by \cite{kurlander24}, who showed that $u$-band observations contribute negligibly to TNO discoveries, whilst $y$-band observations provide only marginal improvements, with no $Y$-band observations of NTs seen within the comparable DES sample \citep{bernardinelli22, bernardinelli23}. Given that NT colors are expected to resemble those of other TNO populations \citep{bolin23, markwardt23}, and that the L4 cloud remains predominantly within the NES region (where only $griz$ observations occur), this simplification has minimal impact on our color characterization results and effectively provides a conservative lower limit on expected yields. We also note that no NT has been discovered with a $g-r\approx0.75$ within the NIRF class, corresponding to the spectral `double-dip' type described in \cite{pinillaalonso25} and surveyed in \cite{markwardt25} — however we again note that with their sample size of eight objects this observed dearth may be due to a lack of observations.

\subsubsection{Orbital Distribution} \label{sec:2.3.3}
Our NT orbital model is based off of those employed in \cite{lin21, alexandersen16, lin16}. Both the eccentricity $e$ and libration amplitude $L_{1:1}$ follow Rayleigh distributions as follows:

\begin{equation}
    p(e) \propto 
        e \cdot \text{exp}\left(- \frac{1}{2} \left( \frac{e}{\sigma_e} \right)^2 \right) de 
\end{equation}

\begin{equation}
    p(L_{1:1}) \propto 
        L_{1:1} \cdot \text{exp}\left(- \frac{1}{2} \left( \frac{L_{1:1}}{\sigma_{L_{1:1}}} \right)^2 \right) dL_{1:1}
\end{equation}
where $\sigma_e / \sigma_{L_{1:1}}$ represent the widths of the respective Rayleigh distributions. For the single-component and rolling power law models, we take these parameters to be the bootstrapped $\sigma_e = 0.045 / \sigma_{L_{1:1}} = 12^\circ$ values from Pan-STARRS1/OSSOS NT observations from \cite{lin21}. \update{For the rolling power law model specifically, we adopt the same single-component dynamical widths because the rolling $H$ distribution from \cite{bernardinelli25} is derived under the assumption of a single underlying population and does not introduce, test for, or provide evidence of distinct hot/cold orbital components within the NTs. Thus, no separate eccentricity distributions are motivated for this case.} For the two-component model, we take the bootstrapped $\sigma_e = 0.02$ for the cold component and $\sigma_e = 0.05$ for the hot component from \cite{lin21} (both in line with other estimates from resonant populations, \citealt{gulbis10, gladman12, pike15, alexandersen16}), and $\sigma_{L_{1:1}} = 12^\circ$ for both. We note that previous studies have shown that for $e > 0.12$ and $L_{1:1} > 35^\circ$ tadpole orbits are no longer stable, and are instead indicative of metastable, or even transient captures \citep{nesvorny09, zhou09, parker15}. As we are only simulating on the timescale of the decade-long survey, and not investigating a purely primordial capture scenario, we choose to probe the full range of $e/L_{1:1}$ to explore the effects of such temporary captures on detections.

The inclination $i$ distribution is modeled as a truncated sin($i$) $\times$ Gaussian distribution \citep{brown01} as follows:

\begin{equation}
    p(i) =
    \begin{cases}
        \sin(i) \cdot \text{exp}\left(- \frac{1}{2} \left( \frac{i}{\sigma_i} \right)^2 \right) \ \ \ \ \ \ \ \ \ \ \ \ i < i_t \\
        0 \ \ \ \ \ \ \ \ \ \ \ \ \ \ \ \ \ \ \ \ \ \ \ \ \ \ \ \ \ \ \ \ \ \ \ \ \ \ \ \ \ \ \ \ \ \ \ \ \ \ \ \ i \geq i_t
    \end{cases}
\end{equation}
where $\sigma_i$ is the width of the Gaussian distribution and $i_t$ is the truncation inclination set at $i_t = 60^\circ$, \update{cf.} \citealt{lin21} and in-line with reasonable survey coverage extent of the LSST. \cite{matheson23, malhotra23, bernardinelli25} have shown a von Mises–Fisher \citep{fisher93} distribution may show a greater statistical significance for modeling the inclination distribution of various TNO dynamical families. However, the concentration parameter $\kappa$ that controls the concentration of the distribution around the mean is poorly measured for the NTs — \cite{bernardinelli25} were not able to satisfactorily constrain the correlation with NIRB/NIRF class using the 6 NTs within the DES sample. As such, we opt to use the simpler Brown distribution instead, as the width $\sigma_i$ was not observed to have any additional color dependency in the analysis of \cite{lin21}. For the single-component and rolling power law models, we again use the bootstrapped inclination width from \cite{lin21} of $\sigma_i=14^\circ$, whilst for the two-component model we have $\sigma_{i,cold}=6^\circ$ and $\sigma_{i,hot}=18^\circ$.
\update{For the rolling power law model, we retain the single-component $\sigma_i$ from \cite{lin21} because the rolling power law $H$ distribution of \cite{bernardinelli25}, whilst containing separate NIRB/NIRF photometric classes, does not identify or require distinct dynamical inclination or eccentricity widths for NTs.}



\begin{figure*}
    \centering
    \includegraphics[width=\textwidth]{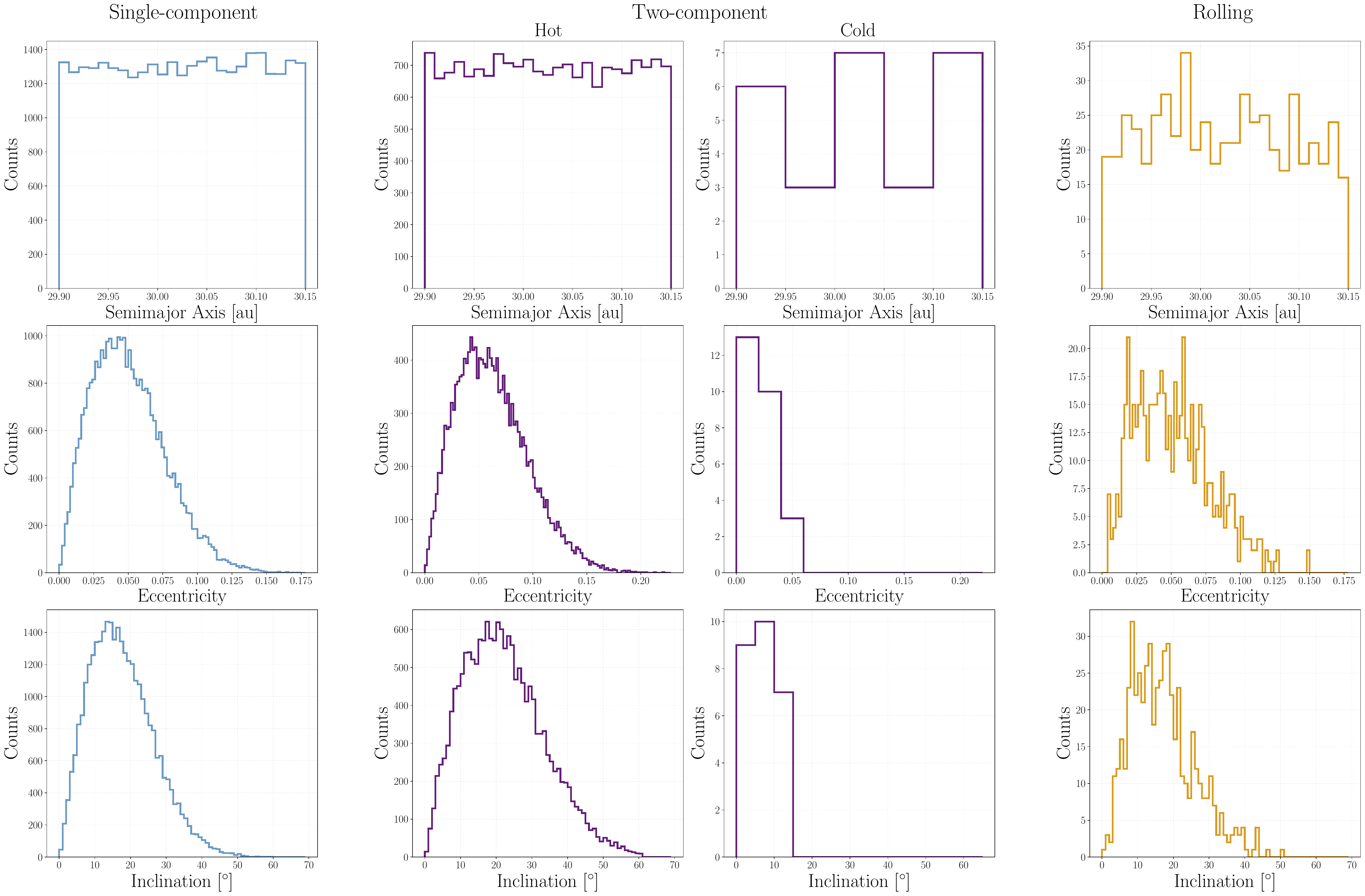}
    \caption{Orbital distributions in (\textit{top row}) semimajor axis $a$, (\textit{middle row}) eccentricity $e$, and (\textit{bottom row}) inclination $i$ for the (\textit{left column}) single-component, (\textit{middle column}) two-component, and (\textit{right column}) rolling power law models respectively.}
    \label{fig:modelorbits}
\end{figure*}

\begin{figure*}
    \centering
    \includegraphics[width=\columnwidth]{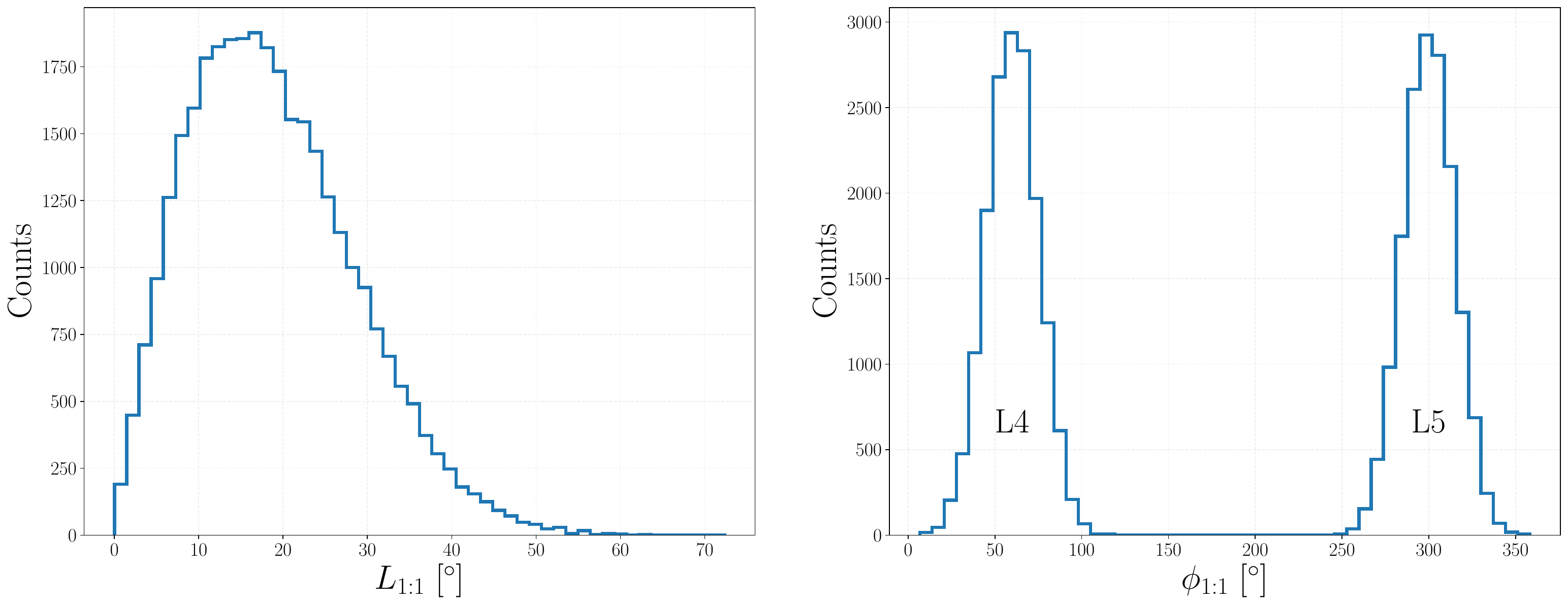}
    \caption{Example distributions of libration angle $L_{1:1}$ and resonance argument $\phi_{1:1}$ used for all 3 $H$ models. All three $H$ models share the same shape of distributions for $L_{1:1}$ and $\phi_{1:1}$, only differing in the number of objects sampled from them.}
    \label{fig:modelangles}
\end{figure*}

For all models, our semimajor axis distribution is uniformly distributed in the range $\mathcal{U}\in$[30.0 au, 30.2 au] - or $\pm0.1$ au from Neptune's semi-major axis value of $\sim$30.1 au (a $<$1\% variation, which \cite{alexandersen16} note that variation on this scale has little effect on NT detectability). \update{The longitude of the ascending node $\Omega$ and mean anomaly $M$ distributions are all sampled from a continuous uniform distribution $\mathcal{U}\in$[0$^\circ$, 360$^\circ$) for all models to account for the effects of orbital precession due to interactions with the giant planets. The resonance argument $\phi_{1:1}$, which controls the split of L4 and L5 clouds, is evenly distributed for all models between the libration centers $\sim$60$^\circ$ and $\sim$300$^\circ$ by the formula:}

\begin{equation} \label{eq:11}
    \phi_{1:1} = \left( \frac{\pi}{3} + R \cdot \frac{2\pi}{3} + L_{1:1} \sin(2\pi \cdot u ) \right) \mod 2\pi
\end{equation}
\update{where $R\in\{-1, 0\}$ and $u$ is uniformly sampled from the continuous distribution $\in[0,1)$. Finally, the argument of perihelion $\omega$ is computed from the resonance condition $\omega = \phi_{1:1} - \Omega - M + \lambda_N$ (where $\lambda_N$ is Neptune's mean longitude), ensuring that $\phi_{1:1}$ is not independent of the orbital angles but still satisfies the required 1:1 resonant geometry.} The resulting orbital distributions in $a/e/i$ for all 3 models are shown in Figure \ref{fig:modelorbits}, with $L_{1:1}/\phi_{1:1}$ shown in Figure \ref{fig:modelangles}.

\section{LSST Predictions} \label{sec:3}

The results presented here are based on the median outcomes of 1000 simulations for each absolute magnitude $H_r$ model described in Section \ref{sec:2.3.1}. For each run, a new NT population model was randomly drawn following the procedure in Section \ref{sec:2.3} and ran through \texttt{Sorcha}. The resulting spread corresponds to uncertainties of $\sim21\%$, $\sim20\%$, and $\sim11\%$ in the discovery yield, illustrated by the shaded regions in Figure \ref{fig:discoveries}. To evaluate the sensitivity of the yield to model assumptions, we also performed additional simulations varying the population scaling parameters (Section \ref{sec:2.3.1}) and the color class fractions $f_{NIRB}$ (Section \ref{sec:2.3.2}). The scaling parameter variations, shown in Appendix \ref{ap:3}, vary the final yields by $^{+82\%}_{-65\%}$, $^{+81\%}_{-63\%}$, and $^{+55\%}_{-55\%}$ for the single-component, two-component, and rolling power law models respectively. By contrast, varying the color class fractions has only a negligible impact, shifting the total yield by at most $\pm\sim3$ objects. For the remainder of this results section, we present results using the best-estimate values for each model parameter.

\subsection{Neptune Trojan Discovery Yield} \label{sec:3.1}
\begin{figure*}
    \centering
    \includegraphics[width=\textwidth]{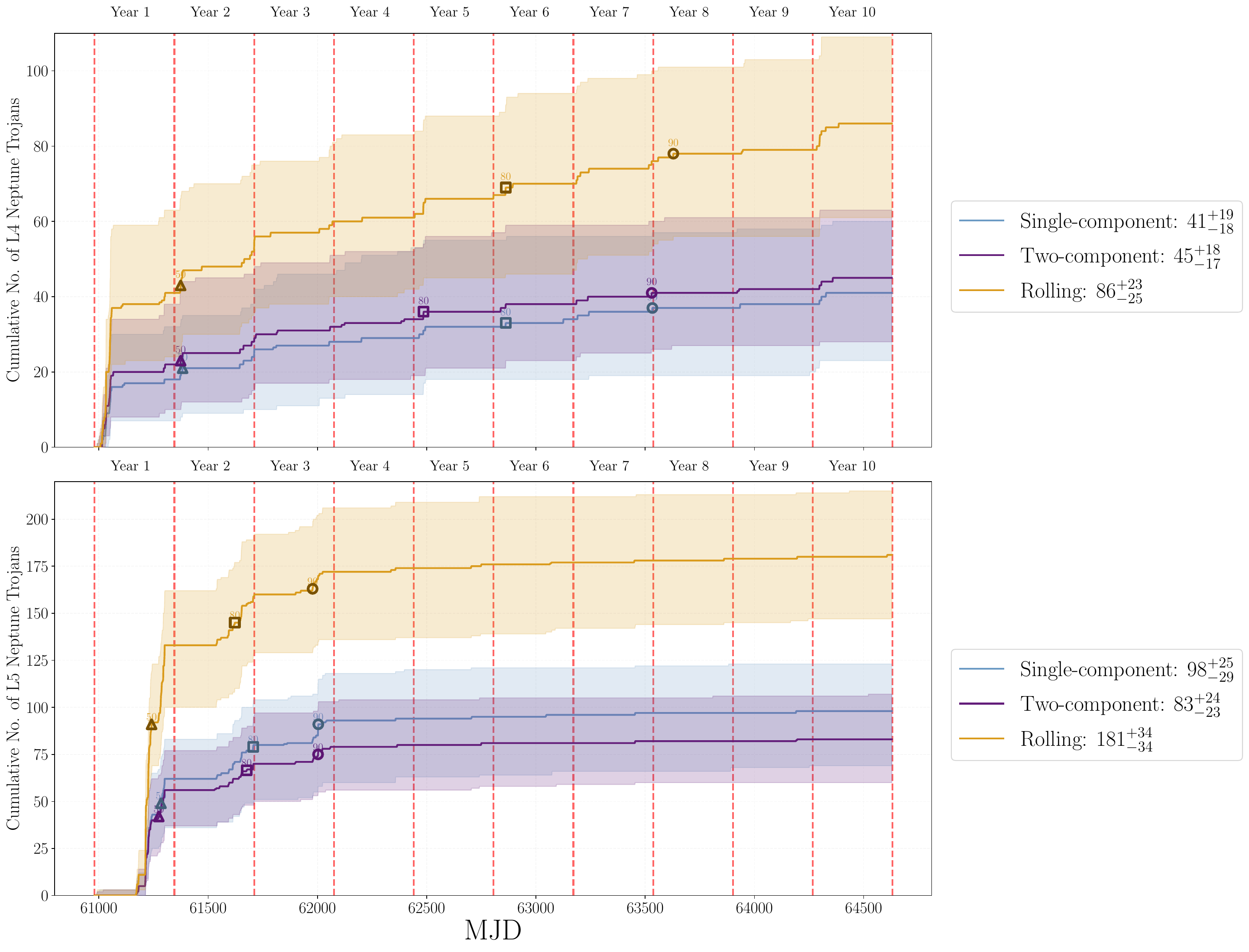}
    \caption{Cumulative histogram (with a bin size of one day) of the discovery rates over the 10 year LSST lifetime for the single-component (blue), two-component (purple), and rolling power law (orange) models. These are further split by (\textit{top row}) L4 discoveries and (\textit{bottom row}) L5 discoveries, with L5 discovery delayed until the cloud is observable in the survey. The points of 50\% (open triangle) 80\% (open square), and 90\% (open circle) completion are shown for each model. Red dashed vertical lines represent survey year start/end points. The pronounced plateau in year 1 discoveries corresponds to an early surge of L4 detections, followed by a lull until sky rotation bring the L5 cloud into an observable airmass.}
    \label{fig:discoveries}
\end{figure*}

The per-Lagrange cloud cumulative ten-year survey discoveries of NTs within the LSST are shown in Figure \ref{fig:discoveries}, with yield totals at 1, 2, 5, and 10 yrs summarized in Table \ref{tab:yield}. Applying a uniform detection efficiency across the whole sky for difference imaging (the true detection efficiency performance of the LSST is still unknown, see Section \ref{sec:3.2} for the effects of the \update{galactic plane}), we find $146^{+34}_{-29}$ single-component, $133^{+28}_{-25}$ two-component, and $285^{+37}_{-30}$ rolling power law NTs discovered. The higher yield in the rolling power law arises from the steeper slope of the bright end of the distribution ($H_r < 8.3$) than the other models, leading to a approximately double the intrinsic number of objects here that the LSST can easily detect. Compared to the 31 NTs known within the MPC, these yields therefore represents a \update{$\sim4-9$ fold} increase in the known population. The current largest individual survey samples of new NT discoveries are the 8 discovered in DES \citep{gerdes16, lin19, bernardinelli20, bernardinelli22}, 6 discovered in the survey of \cite{sheppard10}, 5 discovered in Pan-STARRS1 \citep{lin16}, and the 5 in OSSOS+ \citep{bannister18}. The large leap in NT discoveries compared to prior dedicated NT surveys is owed to the LSST's ability to not just perform relatively narrow, deep searches like prior surveys, but to tile the entire L4 and L5 cloud regions to depths of $m_r \sim 24.7$ \citep{ivezic13, bianco22} multiple times over 10 yrs. This will allow for a more unbiased approach to NT population studies, probing a more complete magnitude and spatial range of the NTs.

\begin{deluxetable}{c c c c c c}[htb]
\tablecaption{Number of Neptune Trojan discoveries by survey year for each absolute magnitude model and split by Lagrange cloud. As a reference, the Minor Planet Center only notes 31 Neptune Trojans, with 27 in L4 and 4 in L5.\label{tab:yield}}
\tablewidth{\columnwidth}
\tablehead{
\multirow{2}{*}{Lagrange} & \multirow{2}{*}{Model} & \multicolumn{4}{c}{LSST Discovery Numbers}  \\ \cline{3-6}
&  & \colhead{1 yr} & \colhead{2 yrs} & \colhead{5 yrs} & \colhead{10 yrs}
}
\startdata
\multirow{3}{*}{L4}    & Single-component & 18  & 24  & 32  & 41 \\
                       & Two-component    & 22  & 28  & 36  & 45 \\   
                       & Rolling          & 41  & 52  & 66  & 86 \\  \cline{1-6}
\multirow{3}{*}{L5}    & Single-component & 62  & 80  & 95  & 98 \\
                       & Two-component    & 56  & 70  & 81  & 83 \\   
                       & Rolling          & 133 & 159  & 176  & 181 \\  \cline{1-6}
\multirow{3}{*}{Combined}    & Single-component & 80  & 104  & 127  & 139 \\
                             & Two-component    & 78  & 98  & 117  & 128 \\   
                             & Rolling          & 174 & 211  & 242  & 267 \\  \cline{1-6}                       
\enddata
\end{deluxetable}

\begin{figure*}
    \centering
    \includegraphics[width=\textwidth]{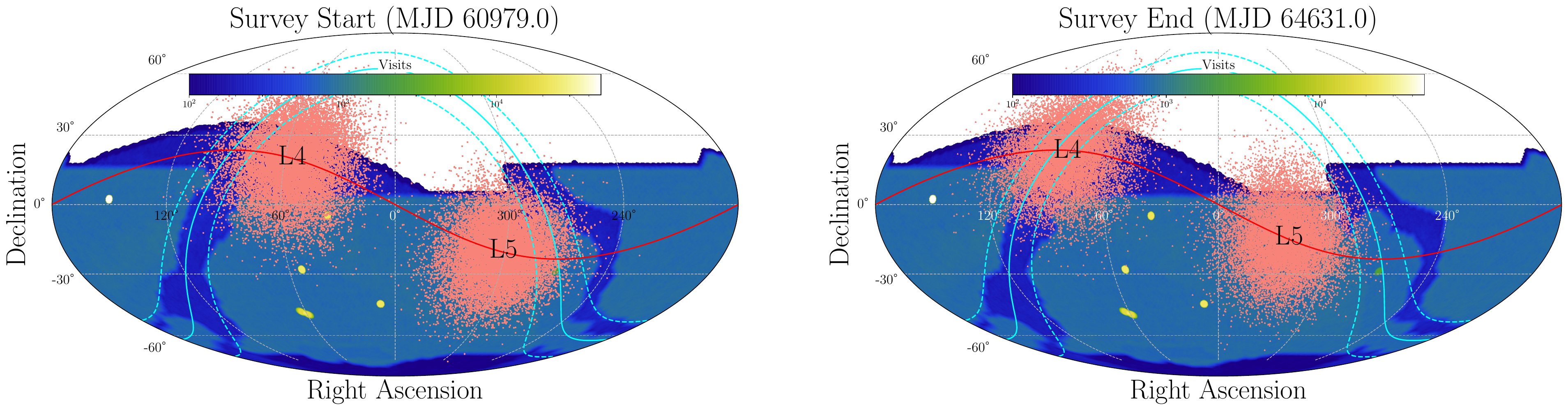}
    \caption{Sky coverage of an entire input intrinsic NT orbital model (pink scatter points) at the beginning of the survey (left) versus at the end of the survey (right) showing the bulk movement of the underlying clouds. In both projections, the L4 cloud is labeled on the left, whereas the L5 cloud is labeled on the right. The red solid line shows the ecliptic plane, and in light blue the \update{galactic plane}, with dashed lines marking $\pm10^\circ$ off of it.}
    \label{fig:cloudmovement}
\end{figure*}

Of each total yield, the discoveries are split in an approximately 1:2 ratio between L4 to L5 cloud NTs respectively — this ratio may change if our initial assumption of symmetric clouds is not accurate. This imbalance arises purely from on-sky localization and survey cadence, and is summarized in Figure \ref{fig:cloudmovement}; from the start of the survey the L4 cloud resides predominantly within the comparatively more sparsely covered NES region (see Figure \ref{fig:cadence}), whilst the L5 cloud remains fully within the WFD footprint over the full survey duration. This is further reflected in Figure \ref{fig:discoveries}, where discoveries of L5 NTs are delayed compared to L4 until the cloud becomes observable within the main survey. Historically, the L5 cloud's proximity to the \update{galactic plane} has made discovery difficult, due to crowded stellar fields hindering difference imaging \cite[e.g.,][]{sheppard10}. The LSST therefore offers an unprecedented opportunity here to systematically measure the L5 cloud in order to constrain the true L4:L5 population ratios, whose (a)symmetry has direct implications for their initial dynamical capture mechanism \citep{sheppard10}. 

Our simulations show that, irrespective of the NT model, the LSST will discover approximately half of its ten-year NT yield within Year 1 (Figure \ref{fig:discoveries}). This rapid early surge is dominated by the brightest objects, with the mean apparent $r$-band magnitude at first detection being $m_r \sim 22.7$. By the second year, this decreases by about one magnitude to \update{an average of} $m_r \sim 23.7$, approaching the LSST's nominal single-visit limiting magnitude \citep[$m_r \sim 24.7$;][]{ivezic13, bianco22}. Beyond this time, discoveries are governed by fainter objects moving inwards and so becoming bright enough to obtain repeat observations that allow them to pass tracklet linkage thresholds for the SSP pipeline. This trend reflects how the LSST's long-term, wide-area coverage enables the eventual recovery of such faint objects, in contrast to prior surveys that were limited to objects bright enough for discovery at time of observation \citep{sheppard10b}.

\begin{figure*}
    \centering
    \includegraphics[width=\columnwidth]{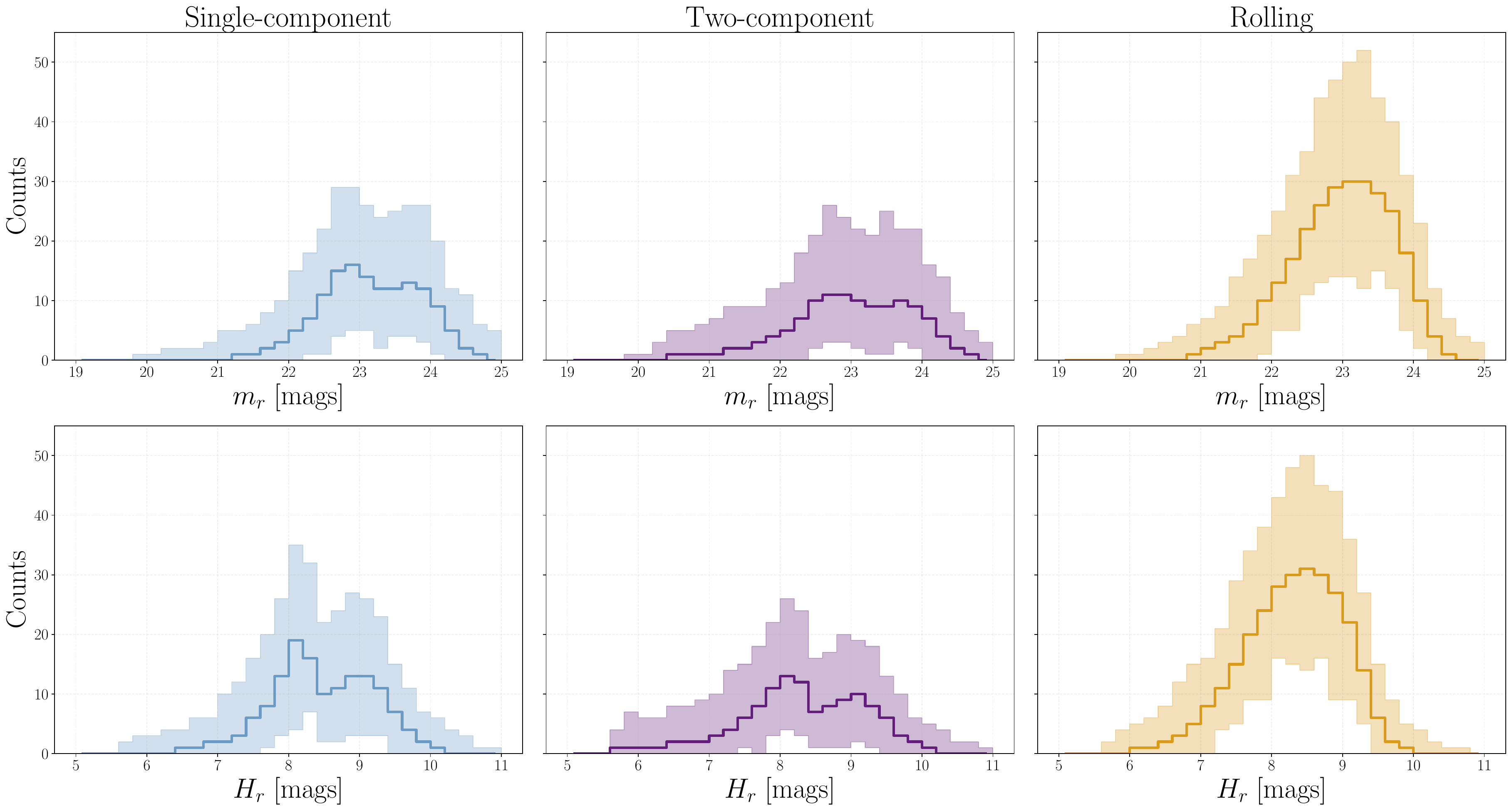}
    \caption{Final ten yr differential histograms of (\textit{top row}) the median apparent magnitude distribution in $r$ band at time of linkage of detected NTs for the three input models, and (\textit{bottom row}) the median absolute magnitude in $r$ band of all detected NTs. The shaded region on each represents the uncertainty on all $10^3$ simulations.}
    \label{fig:detectmags}
\end{figure*}

The final differential distributions of both the apparent and absolute magnitudes at the end of the survey are highlighted in Figure \ref{fig:detectmags}. Detections are cut off at a $H_r = 5.65$, in order to isolate large scattered objects to isolate NTs more likely to have been captured from local primordial material \citep{bernstein04, brown08, petit11, kavelaars21, napier22}. Of the 31 NTs listed in the MPC, only seven have a $H_V \geq 9.0$, with the faintest at $H_V = 10.2$. The LSST NT sample will push this to a depth of $H_V\approx10.2-11.2$ \citep[assuming $V-r=0.2$, \update{cf.} Jupiter Trojans,][]{lin21}, probing NT sizes down to $\sim15$ km \citep[assuming a 5\% albedo, again \update{cf.} Jupiter Trojans,][]{fernandez03, fernandez09, sheppard10b}. More importantly, we predict the LSST will discover $\sim70-130$ NTs brighter than the break magnitude $H_B=8.3$, and $\sim60-150$ fainter, compared to the 13 and 18 known in the MPC respectively. The LSST will thus ensure coverage well into both the large and small regimes of the size-frequency distribution; bright/large NT statistics, where the slope reflects the primordial population size distribution, can be used to directly compare NT capture mechanism models \citep{morbidelli09}, whilst the faint/small NT end will allow for probing commensuration with other dynamically hot TNO families \citep{fraser14}. Crucially, this $H$ range spans a region sensitive to long-term collisional evolution \citep{bottke23}, which will assist in discrimination between a slope shaped by direct capture of NTs versus one modified by Gyr-scale collisional grinding.

\subsection{Galactic Plane Observations} \label{sec:3.2}
The high stellar densities within $\pm10^\circ$ of the \update{galactic plane} \citep{gilmore83, chen01, juric08, valenti16} can make NT discovery more difficult due to overlapping and blending of the star/NT point-spread functions, and the resulting challenge of generating reliable sky templates that accurately subtract stellar backgrounds for difference imaging \citep{olsen03, kubica07}. Whilst our baseline simulations assume perfect difference imaging capability, real observations within the \update{galactic plane} will inevitably suffer reduced single-visit efficiency. To quantify the impact, we removed all observations within $\pm10^\circ$ of the \update{galactic plane} from the v4.3.1 cadence simulation, and re-ran our models. This excision only reduced the total NT yield by $\sim15\%$, and the median number of observations per object by $\sim8\%$ for all models, providing a lower limit on NT yields — this minimal impact is largely because the L5 cloud moves away from the plane already over the survey duration, and the L4 cloud, residing primarily in the NES region, receives less visits overall than the L5 within the WFD region (see Figure \ref{fig:cloudmovement}). We, however, also provide a more nuanced approach to NT detection by applying an empirical detection efficiency penalty derived from the search for NTs in high stellar density fields by \cite{sheppard10}. This produces an even smaller effect, reducing the initial cadence scenario by $<1\%$ and instead slightly shifting discoveries to periods of survey cadence not within the plane. These results indicate that the \update{galactic plane} has minimal long-term impact on the LSST's NT yield — bright objects in the plane still accumulate sufficient detections over the 10 yr survey to meet linking thresholds, even if single-visit performance is temporarily degraded.

\subsection{Observation Numbers and Orbit Arcs} \label{sec:3.3}
Over the 10 yr baseline, each detected NT in our simulations accumulate $\sim240-290$ observations per object, excluding DDF visits — this assumes that the SSP linking behavior performs to expected efficiency \citep{juric20}, and that precovery (i.e., recovering observations of the object post-linkage) is possible. From \update{LSST} observations \update{alone}, $>95\%$ are observed for approaching the full survey duration, giving them orbital arcs $>9$ yr. The long, dense arcs that the LSST \update{by itself} can provide here will be more than adequate at lowering orbital uncertainties enough to distinguish between the 1:1 resonance from other nearby ones (or other scattering populations), and allow for accurate long-term orbit integrations to calculate libration amplitude to classify NTs as stable or metastable within the resonance. In our sample, the contamination from potentially transient or metastable objects is small, with $\sim$10 NTs with $e > 0.12 / L_{1:1} > 35^\circ$, indicating that these should not affect primordial population studies significantly.

Due to the spatial extents of the L4 and L5 clouds, entrance into any of the DDF areas is exceedingly rare — over $10^3$ simulations, on average no NTs ever enter either the COSMOS ($150^\text{h}06^\text{m}36^\text{s}/+02^\text{d}13^\text{m}48^\text{s}$), or the X-ray Multi-Mirror Mission Large Scale Structure (XMM-LSS, or XMM; $35^\text{h}34^\text{m}12^\text{s}/-04^\text{d}49^\text{m}12^\text{s}$), with a maximum of three L4 NTs within XMM in any simulation. Those few L4 NTs that do pass through XMM remain for about a year, and receive at least twice the number of total observations across all filters as a WFD-only NT. Population-wide NT studies will thus come from WFD discoveries — however if any do enter XMM, with $\sim250$ observations for single- and two-component NTs and $\sim290$ for rolling power law NTs, they will allow for dense light curve coverage and detailed case-studies on photometric variability.

\subsection{Light Curves and Colors}
\begin{figure}
    \centering
    \includegraphics[width=\linewidth]{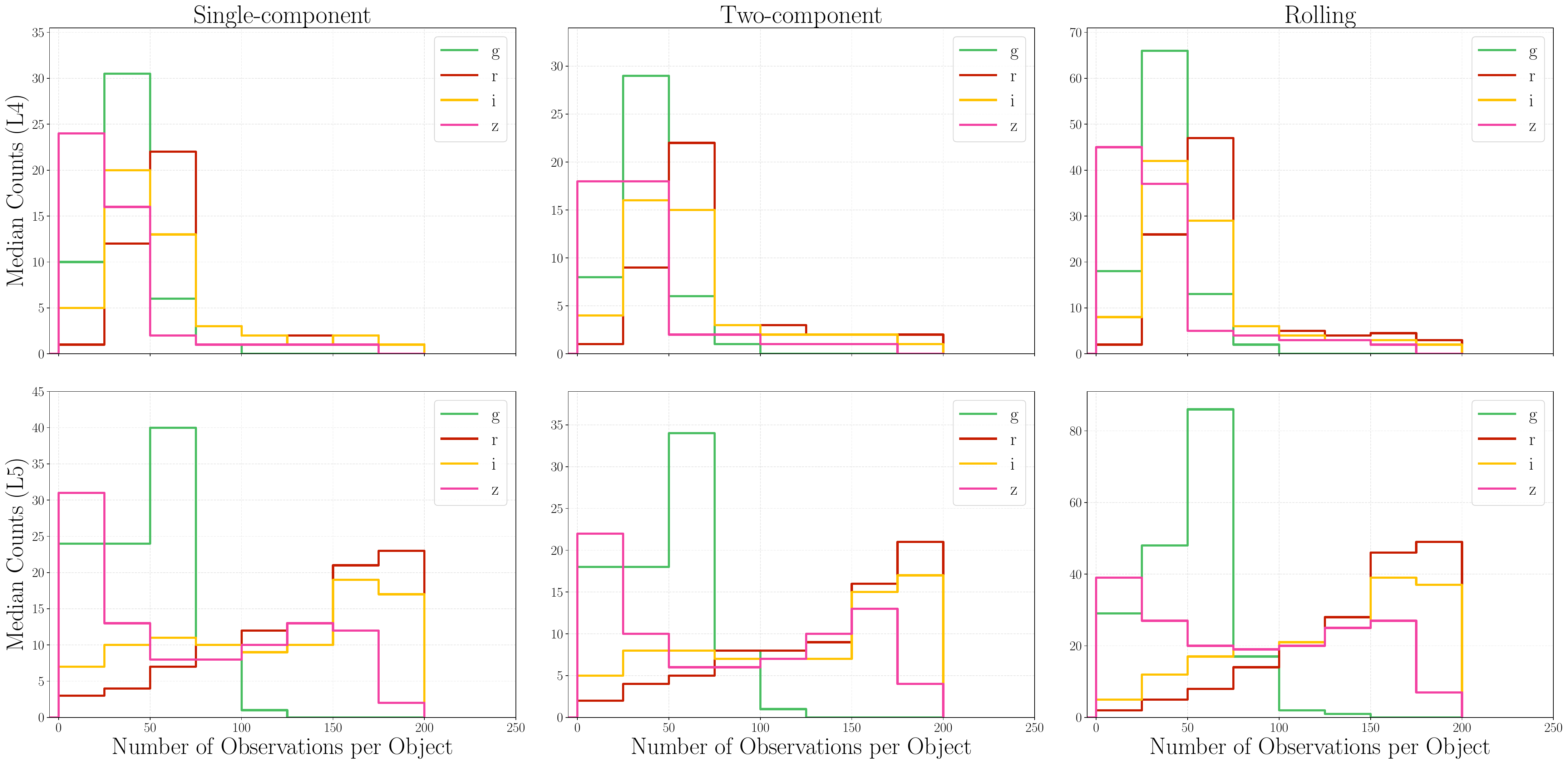}
    \caption{Histograms of the number of detections per object, broken down per filter for the three input models (columns) and by Lagrange cloud (L4 top row, L5 bottom row). All exclude any observations that fall within a DDF.}
    \label{fig:colobs}
\end{figure}

The LSST's cadence will also provide rich per-object coverage across all filters, with the distribution in visits per filter, separated by Lagrange cloud, shown in Figure \ref{fig:colobs}, and summarized in Table \ref{tab:colcounts} — the bimodality in $r/i/z$-band observations that is not seen in $g$ is a reflection on the input NT model being intrinsically less bright in $g$, combined with the WFD L5 NTs being observed $\sim2-3$ more times on average than NES L4 NTs, producing bimodality in the brighter $r/i/z$ bands. The average signal-to-noise S/N per object is at $\sim40-70$, or a magnitude uncertainty of $\sim0.015-0.025$ mags. With average observations of 15-30 per object per year, and given the ability to observe both clouds, the LSST will be able to begin the first comprehensive NT light curve study — probing the similarity of photometric variability relative to the wider TNO population to better constrain their origin \citep{jewitt18}, and providing an expanded color sample from prior surveys \citep{sheppard06, bolin23, markwardt23}. Whilst the survey cadence and use of multiple broadband filters is particularly excellent for long-term monitoring and color sampling of objects, the long inter-night visit gaps per filter \citep[$\sim$week(s),][]{jones14, bianco22} may limit rotation sensitivity to longer periods — nevertheless a broad population-level study will still be possible.

In order to investigate the number of NTs that will have a well-measured color light curve, we apply the quality metrics as defined in \cite{schwamb23}, as derived from the LSST Metrics Analysis Framework \citep[MAF,][]{jones14}. These metrics require that, for a given object, at least one band (primary) has $>30$ observations, and all other (secondary) bands must have $>20$ — all with S/N $> 5$. Physically, this is meant to provide a proxy to fitting a light curve in the primary band and utilize the secondary to refine the fit and improve color measurements — for objects with less photometric variability these metrics however not be feasible. Applying these metrics to our dataset yields $\sim60\%$ of NTs ($\approx87, 80,$ and $171$ for the single-component, two-component, and rolling power law models respectively) passing the metrics — a slight increase on the TNOs from \cite{kurlander25}, but broadly attributable to our model containing proportionally more relatively bright objects. This sample, on the lowest estimate, more than quadruples the sample size used in the surveys from \cite{bolin23} (18), \cite{markwardt23} (15), and \cite{jewitt18} (13). An quadrupled sample size with precisely measured surface colors from phase curve fitting will allow the LSST to more accurately measure the exact red/very-red color fraction \citep{bolin23}, and directly test the tentative absolute magnitude-color correlation reported by \cite{markwardt23}. These objects will also supply many well-characterized targets for spectroscopic follow-up to probe red/very-red compositional differences \citep{markwardt25}.

\begin{deluxetable}{c c c c c c c}
\tablecaption{Median number of detections of NTs, \update{for all observations irrespective of filter, and} split across filters over the 10 year LSST survey duration (excluding DDF visits).\label{tab:colcounts}}
\tablewidth{\columnwidth}
\tablehead{
\multirow{3}{*}{Lagrange} &
\multirow{3}{*}{Model} &
\multicolumn{5}{c}{Median number of} \\ 
& & \multicolumn{5}{c}{observations per object} \\ \cline{3-7}
& & \colhead{\update{All Filters}} & \colhead{g} & \colhead{r} & \colhead{i} & \colhead{z} 
}
\startdata
\multirow{3}{*}{L4} & Single-component & 165 & 36  & 55   & 48  & 25 \\
                    & Two-component    & 176 & 37  & 57   & 52  & 29 \\   
                    & Rolling          & 169 & 36  & 55   & 49  & 26 \\ \cline{1-7}
\multirow{3}{*}{L5} & Single-component & 389 & 50  & 148  & 127  & 64 \\
                    & Two-component    & 427 & 54  & 156  & 139  & 79 \\   
                    & Rolling          & 440 & 56  & 159  & 143  & 83                    
\enddata
\end{deluxetable}

\section{SV Survey Predictions} \label{sec:4}

Whilst the main survey will ultimately dominate NT discoveries, the interim Science Validation (SV) survey (see Section \ref{sec:2.2}) will offer a unique preview of the LSST's early NT yield. Simulating observations with the \update{\texttt{lsstcam\_20250930}} cadence, and assuming perfect template generation for difference imaging, we predict median discovery yields over $10^3$ simulations of \update{$33^{+18}_{-15}$, $32^{+14}_{-16}$, and $72^{+34}_{-21}$} for the single-component, two-component, and rolling power law models respectively. The survey's limited coverage and cadence is overwhelmingly biased towards observing the L5 cloud (see Figure \ref{fig:cloudmovement}), with $<2\%$ of discoveries coming from the L4 cloud. Across these simulated detections, individual objects receive \update{a median of $\sim30$} visits, broken down in \update{4 $g$, 10 $r$, 8 $i$, 6 $y$} observations \update{per filter}, providing a rich dataset for preliminary photometric variability and color measurements. \update{All simulated detections occur before 2025 September 21, the end of the real observations included in the final SV database. This means the simulated discovery samples correspond entirely to epochs already captured on sky, implying that the SV dataset may already contain counterparts to many of the NTs predicted here. These represent potential viable NT candidates awaiting identification in the existing images within Data Preview 2 \citep{guy21}.} 


\section{Conclusions} \label{sec:5}

In this work, we have combined the current best understanding of the Neptune Trojan orbital and surface color models together with the survey simulator \texttt{Sorcha} and a cadence simulation of the LSST in order to produce the very first estimates for the observability of the NTs in the LSST. With the LSST's unprecedented depth, cadence, and sky coverage we find that there will be a new era of NT science, including:

\begin{itemize}
    \item The predicted number of NT discoveries within the LSST is 146, 133, and 285 for a single-component, two-component, and rolling power law model absolute magnitude distributions, respectively. This represents a $\sim4-9$ fold increase on the known 31 NTs within the MPC dataset today.
    \item There is a bias towards discovery and observation of L5 NTs over L4 NTs, despite an intrinsically symmetric cloud distribution, due to the L4 residing predominantly within the lower cadence NES region of the LSST survey footprint. The L5 cloud has historically been understudied however, due to its proximity to the high stellar density \update{galactic plane}, so the LSST will here offer the first opportunity to systematically investigate any (a)symmetries between the two.
    \item Removing \update{galactic plane} observations altogether from our simulations only reduces the yield of each model by $\sim15\%$, approximately the same level as the uncertainty due to simulation randomization. Objects which are bright enough will still accumulate sufficiently high S/N observations over the ten year survey to pass linking thresholds regardless (assuming idealized SSP performance).
    \item Nearly all of each NT model ($>95\%$) will achieve orbital arcs $>9$ yr, allowing for well constrained orbit measurements to confirm stability within the 1:1 resonance
    \item $60\%$ of each NT model pass quality color light curve metrics, giving at least a fourfold increase on the color samples used in previous color studies, allowing for more robust tests of absolute magnitude-color correlations and red/very-red color population fractions.
\end{itemize}

While the results presented in this work provide a detailed forecast for NT detections in the LSST era, they remain dependent on several key assumptions that warrant caution. Chief among these is the small sample size from which our input models are drawn: current constraints on the absolute magnitude distributions are based on fewer than 30 known objects, most of which reside in the L4 cloud \citep{lin21}. The underlying population parameters, particularly size distributions, therefore carry significant uncertainties. We have quantified this through running 1000 simulations, yielding an estimated uncertainty of $\pm 10-23\%$ due to simulation randomization, $^{+82\%}_{-65\%}$, $^{+81\%}_{-63\%}$, and $^{+55\%}_{-55\%}$ from absolute magnitude scaling uncertainty for the single-component, two-component, and rolling power law models respectively (see Appendix \ref{ap:3}), and $\pm 1-2\%$ from color uncertainty. Yet even under the most conservative yield estimate in the two component model, the known NT population size is set to quadruple from the known MPC population.

Further caveats include assumptions made regarding the template generation process. Our simulation assumes static, pre-generated templates are available for all sky regions prior to survey commencement, which does not reflect reality. \cite{schwamb23} and \cite{robinson25} discuss in detail the impact that such a lack of template generation will have on year 1 statistics for solar system objects — whilst real-time discoveries may vary due to this, observations will still be able to be recovered as incremental template generation proceeds and data is reprocessed for Data Release 1 \citep{guy21}. Our modeled detection efficiency also presumes idealized performance from the Solar System Processing pipelines - in practice however, linking efficiency, false positive rates, and tracklet fitting performance may reduce the effective yield, particularly in lower-cadence regions like the NES, or high stellar background regions like the \update{galactic plane}. Finally, the precise commencement date of observing with the LSST will also have an effect on observations numbers due to continuous movement of the clouds placing them in better or worse alignment with regions like the DDFs. If the LSST were to start observations later than November 2025, this will result in further drifts of both clouds, affecting their observability and potentially altering our yield estimates.  

Despite these uncertainties, our results reinforce the exception potential for the LSST to revolutionize NT science. With detections extending to magnitudes approaching $m_r \sim 24$, and arc lengths approximately a decade for most objects, the LSST will enable detailed orbital classification, population modeling, and physical characterization - particularly for the underexplored L5 cloud. The breadth and depth of data anticipated here will mark the beginning of a statistically robust era in outer solar system dynamics, capable of addressing longstanding questions on the origin and evolution of this 1:1 resonant population.

\begin{acknowledgments}
J.Murtagh acknowledges support from the Department for the Economy (DfE) Northern Ireland postgraduate studentship scheme and travel support from the STFC for UK participation in LSST through grant ST/S006206/1. J.Murtagh and J.A.K. thank the LSST-DA Data Science Fellowship Program, which is funded by LSST-DA, the Brinson Foundation, and the Moore Foundation; their participation in the program has benefited this work. M.E.S. acknowledges support in part from UK Science and Technology Facilities Council (STFC) grants ST/V000691/1 and ST/X001253/1. P.H.B., J.A.K., M.J., P.Y., J.Moeyens and C.O.C. acknowledge the support from the University of Washington College of Arts and Sciences, Department of Astronomy, and the DiRAC (Data-intensive Research in Astrophysics and Cosmology) Institute. The DiRAC Institute is supported through generous gifts from the Charles and Lisa Simonyi Fund for Arts and Sciences, Janet and Lloyd Frink, and the Washington Research Foundation.  M.J.H. gratefully acknowledges support from the NSF (grant No. AST2206194) and the NASA YORPD Program (grant No. 80NSSC22K0239).

This research has made use of NASA's Astrophysics Data System Bibliographic Services. This research has made use of data and/or services provided by the International Astronomical Union's Minor Planet Center. The SPICE Resource files used in this work are described in \citet{acton96, acton18}. Simulations in this paper made use of the REBOUND N-body code \citep{rein12}. The simulations were integrated using IAS15, a 15th order Gauss-Radau integrator \citep{rein15}. Some of the results in this paper have been derived using the \texttt{healpy} and HEALPix packages. This work made use of Astropy:\footnote{http://www.astropy.org} a community-developed core \python package and an ecosystem of tools and resources for astronomy \citep{astropy13,astropy18,astropy22}. We thank the Vera C. Rubin Observatory Data Management Team and Scheduler Team for making their software open-source. This research received
support through Schmidt Sciences.

We are grateful for the use of the computing resources from the Northern Ireland High Performance Computing (NI-HPC) service funded by EPSRC (EP/T022175). 

\update{We thank the anonymous reviewer for their helpful comments, their insight has improved this manuscript.}

\end{acknowledgments}

\begin{contribution}

J.Murtagh was responsible for writing and submitting the manuscript. He ran all of the simulations and performed all of the analysis detailed in this manuscript. 

M.E.S supervised and contributed to the overall discussion and conception of the initial research concept. She provided meaningful edits and proofreads to this manuscript. 

P.H.B. contributed to the development and design of the models used within this work, including provision of original curated data and code. He also provided meaningful edits and proofreads of this manuscript.

H.W.L. contributed to the initial research concept and provided original curated code to develop the models used within this work. He also provided meaningful edits and proofreads of this manuscript.

J.A.K. provided the initial plotting scripts to produce skymaps in Figures \ref{fig:cadence} and \ref{fig:cloudmovement}. 

G.F. and M.J.H. provided meaningful edits and proofreads of the manuscript and contributed significantly to the development of the survey simulator \texttt{Sorcha} used in this work.

R.L.J., P.Y. produced the cadence simulations of the Rubin Observatory used within this work.

S.R.M., S.C., M.J., S.E., J.Moeyens, J.K., D.O., M.W., and C.O.C. all contributed significantly to the development of the survey simulator \texttt{Sorcha} used in this work. 


\end{contribution}

%

\software{Sorcha \citep{merritt25, holman25}, 
          ASSIST \citep{holman23, rein23}, 
          Astropy \citep{astropy13, astropy18}, 
          CMasher \citep{vandervelden20},
          Healpy \citep{zonca19, gorski05}, 
          Jupyter Notebook \citep{kluyver16}
          Matplotlib \citep{hunter07},
          Numba \citep{lam15}, 
          Numpy \citep{harris20}, 
          Pandas \citep{mckinney10, pandas20}, 
          Pooch \citep{uieda20}, 
          PyTables \citep{pytables02}, 
          REBOUND \citep{rein12, rein15}, 
          rubin\_sim \citep{bianco22, yoachim23}, 
          rubin\_scheduler \citep{naghib19, yoachim24b}, 
          sbpy \citep{mommert19}, 
          SciPy \citep{virtanen20},
          seaborn \citep{waskom21},
          spacerocks \citep{napier20},
          Spiceypy \citep{annex20}, 
          sqlite (\href{https://www.sqlite.org/index.html}{https://www.sqlite.org/index.html}), 
          sqlite3 (\href{https://docs.python.org/3/library/sqlite3.html}{https://docs.python.org/3/library/sqlite3.html}), 
          tqdm \citep{dacostaluis23}
          }

\appendix
\restartappendixnumbering

\section{Deriving the Divot Power Law} \label{ap:1}

\begin{figure}
    \centering
    \includegraphics[width=\columnwidth]{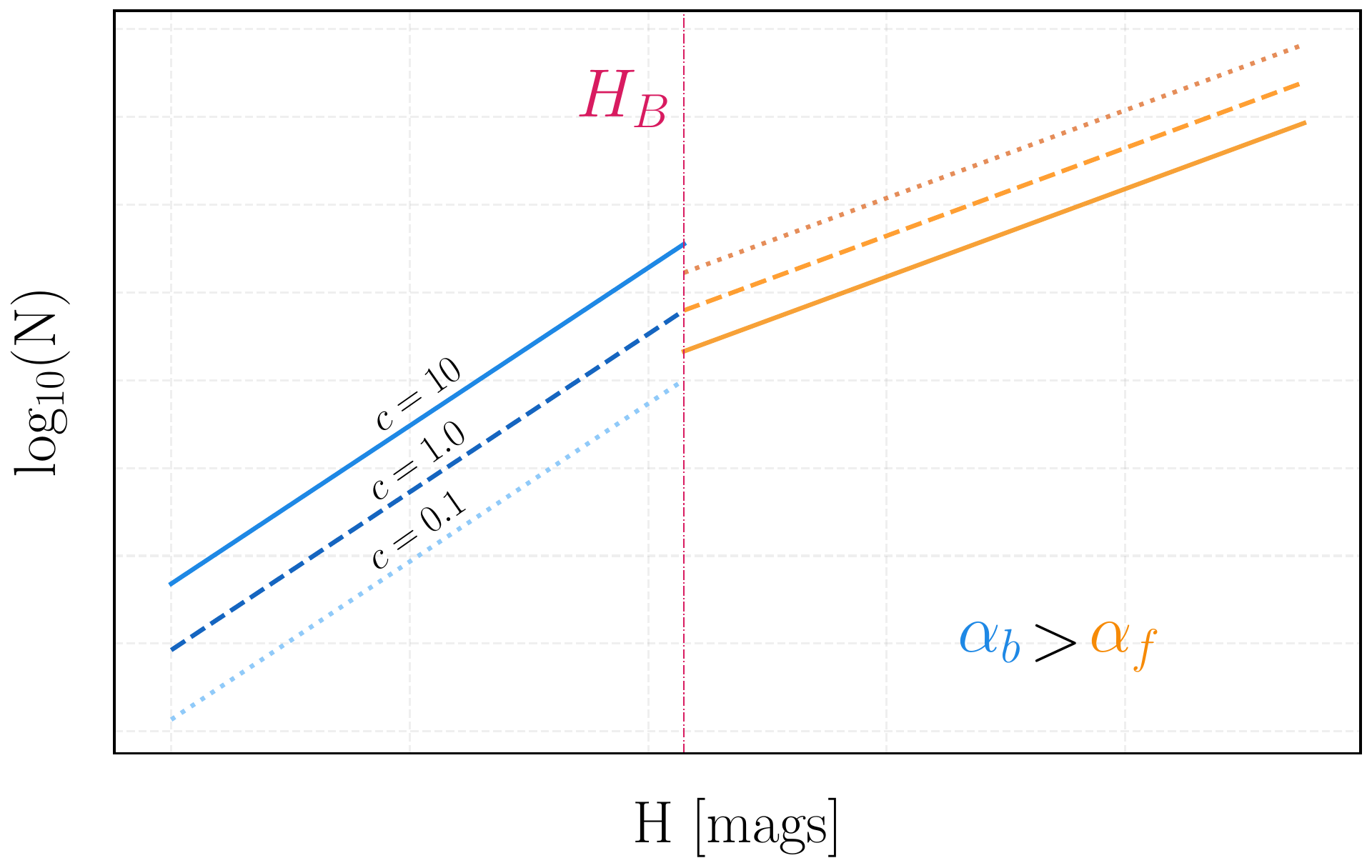}
    \caption{Graphical representation of the differential power law. The `bright' end of the distribution (blue) is quantified by a slope $\alpha_b$, whereas the `faint' end (orange) is quantified by a slope $\alpha_f$ (in this plot, $\alpha_b > \alpha_f)$, with a break at $H_B$. The size of the divot at $H_B$ is quantified by a contrast parameter $c$, that is the ratio of the differential binned objects pre-divot to the differential binned objects post-divot. For $c = 1.0$ (dashed), the power law reduces to a broken `knee' form, whereas for $c<1$ and $c>1$, a `divot' power law occurs, either increasing (dotted) or decreasing (solid) the number of objects post-break respectively.}
    \label{fig:divotlaw}
\end{figure}

A broken (or ``knee'') power law describes a distribution whose slope changes abruptly at some break value, often used to model size-frequency distributions of small bodies. A divot variant additionally includes a multiplicative drop in number density at some break value, producing a discontinuity between the slopes either side. From \cite{shankman12}, the differential form of the `divot' power law in an infinitesimal absolute magnitude $H$ range is given by:

\begin{subnumcases}{\frac{dN}{dH} = }
    \begin{array}{l} k_b \alpha_b \ln(10) 10^{\alpha_b H} \\ \ \ \ \left( \text{where} \ \ k_b = A \cdot 10^{-\alpha_b H_0} \right)  \end{array} &   $H < H_B$       \label{ap:eq1a} \\
    \begin{array}{l} k_f \alpha_f \ln(10) 10^{\alpha_f H} \\ \ \ \ \left( \text{where} \ \ k_f = B \cdot 10^{-\alpha_f H_B} \right)  \end{array}  &   $H \geq H_B$    \label{ap:eq1b}
\end{subnumcases}
where $k_b / k_f$ are normalization parameters (with constants $A/B$) for the `bright' and `faint' ends of the distribution with respect to an observationally determined normalization absolute magnitude $H_0$ and the break absolute magnitude $H_B$ respectively. $\alpha_b / \alpha_f$ are the slope parameters of the bright and faint ends of the distribution. This is shown diagrammatically in Figure \ref{fig:divotlaw}. The cumulative total number of objects after the break magnitude $N(<H_B)$ is found by the sum of the integrals of the bright and faint components of the distribution over the appropriate pre- and post-break magnitude ranges respectively. Integrating first Equation \ref{ap:eq1a}:

\begin{equation} \label{ap:eq2}
    \begin{aligned}
        N(<H_B) & = \int_{-\infty}^{H_B} dN \\
                & = \int_{-\infty}^{H_B} k_b \alpha_b \ln(10) 10^{\alpha_b H} dH \\
                & = k_b \alpha_b \ln{10} \frac{1}{\alpha_b \ln(10)} \left. 10^{\alpha_b H} \right|_{\infty}^{H_B} \\
                & = k_b \left[ 10^{\alpha_b H_B} - \cancelto{=0}{10^{-\infty}} \right] \\
                & = k_b 10^{\alpha_b H_B}
    \end{aligned}
\end{equation}

Now integrating the `faint' end of the distribution:

\begin{equation} \label{ap:eq3}
    \begin{aligned}
        N(H_B \leq H \leq H') & = \int_{H_B}^{H'} dH \\
                           & = \int_{H_B}^{H'} k_f \alpha_f \ln(10) 10^{\alpha_f H} dH \\
                           & = k_f \alpha_f \ln(10) \frac{1}{\alpha_f \ln(10)} \left. 10^{\alpha_f H} \right| \\
                           & = k_f \left[ 10^{\alpha_f H'} - 10^{\alpha_f H_B} \right]
    \end{aligned}
\end{equation}

Both components are matched at the divot according to a ratio of the infinitesimal number of objects before the divot to after the divot, quantified by the contrast parameter $c$. The faint end normalization parameter $k_f$ can therefore be rewritten in terms of the bright end normalization parameter $k_b$ as follows:

\begin{equation} \label{ap:eq4}
    \begin{aligned}
        N_{bright}(H=H_B) & = c \cdot N_{faint}(H = H_B) \\
        k_b \alpha_b \ln(10) 10^{\alpha_b H_B}  & = c \cdot k_f \alpha_f \ln(10) 10^{\alpha_f H_B} \\
        k_f & = k_b \frac{1}{c} \frac{\alpha_b}{\alpha_f} \frac{10^{\alpha_b H_B}}{10^{\alpha_f H_B}}
    \end{aligned}
\end{equation}

Inserting this expression of $k_f$ in to Equation \ref{ap:eq3} then,

\begin{equation} \label{ap:eq5}
    \begin{aligned}
        N(H_B \leq H \leq H') = k_b \frac{1}{c} \frac{\alpha_b}{\alpha_f} \frac{10^{\alpha_b H_B}}{10^{\alpha_f H_B}} \left[ 10^{\alpha_f H'} - 10^{\alpha_f H_B} \right]
    \end{aligned}
\end{equation}

Therefore, the cumulative total number of objects after the break with an arbitrary $H=H'$ is given as the sum of Equation \ref{ap:eq2} and Equation \ref{ap:eq5}:

\begin{equation} \label{ap:eq6}
    N({\leq}H') = \underbrace{k_b 10^{\alpha_b H_B}}_{H < H_B} + \underbrace{k_b \frac{1}{c} \frac{\alpha_b}{\alpha_f} \frac{10^{\alpha_b H_B}}{10^{\alpha_f H_B}} \left[ 10^{\alpha_f H'} - 10^{\alpha_f H_B} \right]}_{H \geq H_B}
\end{equation}
If the measured $H_0 \geq H_B$, the value of $k_B$ can be determined by rearranging Equation \ref{ap:eq6} and solving using the known $N(H < H_0)$ - otherwise, if $H_0 < H_B$, the value of $k_b$ must be determined by solving the bright end distribution, Equation \ref{ap:eq2}.

Once the normalization parameter $k_b$ is determined, in order to uniformly draw from this distribution, one draws from the two distributions over their respective $H$ ranges such that they contain an appropriate fraction of objects described simplifying Equation \ref{ap:eq5} by factorizing Equation \ref{ap:eq2} out as follows:

\begin{equation} \label{ap:eq7}
    \begin{aligned}
        N({\leq}H') = N_{total} & = k_b 10^{\alpha_b H_B} + k_b \frac{1}{c} \frac{\alpha_b}{\alpha_f} \frac{10^{\alpha_b H_B}}{10^{\alpha_f H_B}} \left[ 10^{\alpha_f H'} - 10^{\alpha_f H_B} \right] \\
                                & = k_b 10^{\alpha_b H_B} + k_b 10^{\alpha_b H_B} \frac{1}{c} \frac{\alpha_b}{\alpha_f} \left[ 10^{\alpha_f (H' - H_B)} - \cancelto{=1}{10^{\alpha_f (H _B -H_B)}} \right] \\
                                & = N_b \left( 1 + \frac{1}{c} \frac{\alpha_b}{\alpha_f} \left( 10^{\alpha_f (H' - H_B)} - 1 \right) \right) \\
        \frac{N_b}{N_{total}}   & = \left( 1 + \frac{1}{c} \frac{\alpha_b}{\alpha_f} \left( 10^{\alpha_f (H' - H_B)} - 1 \right) \right)^{-1}
    \end{aligned}
\end{equation}

In order to uniformly draw from each distribution, one creates an inverse transform of the cumulative distribution functions described Equations \ref{ap:eq2} and \ref{ap:eq5} such that $N(<H)^{-1}(\mathcal{U})$, where $\mathcal{U}$ is uniformly distributed in the range $\in [0,1]$. Taking the bright end first,

\begin{equation} \label{ap:eq8}
    \begin{aligned}
        F(H) & = \frac{\cancel{k_b \alpha_b \ln(10)} \int_{-\infty}^{H'} 10^{\alpha_b H} dH}{\cancel{k_b \alpha_b \ln(10)} \int_{-\infty}^{H_B} 10^{\alpha_b H} dH} \\
             & = \frac{10^{\alpha_b H'} - \cancelto{=0}{10^{-\infty}}}{10^{\alpha_b H_B} - \cancelto{=0}{10^{-\infty}}} \\
             & = \begin{aligned}[t] \frac{10^{\alpha_b H'}}{10^{\alpha_b H_B}} & = \mathcal{U} \\
                                                              10^{\alpha_b H'} & = \mathcal{U} \cdot 10^{\alpha_b H_B} \\
                                                                   \alpha_b H' & = \log_{10}(\mathcal{U}) + \alpha_b H_B \\
                                                                            H' & = \frac{1}{\alpha_b} \left( \log_{10}(\mathcal{U}) + \alpha_b H_B \right) \\
            \end{aligned}
    \end{aligned}
\end{equation}

Now inverting the faint end:

\begin{equation} \label{ap:eq9}
    \begin{aligned}
        F(H) & = \frac{\cancel{k_f \alpha_f \ln(10)} \int_{H_B}^{H'} 10^{\alpha_f H} dH}{\cancel{k_f \alpha_f \ln(10)} \int_{H_B}^{H_{max}}  10^{\alpha_f H} dH} \\
             & = \begin{aligned}[t] \frac{10^{\alpha_f H'} - 10^{\alpha_f H_B}}{10^{\alpha_f H_{max}} - 10^{\alpha_f H_B}} & = \mathcal{U} \\
                                                                                                          10^{\alpha_f H'} & = \mathcal{U} \cdot \left[ 10^{\alpha_f H_{max}} - 10^{\alpha_f H_B} \right] + 10^{\alpha_f H_B} \\
                                                                                                            \alpha_f H' & = \log_{10}(\mathcal{U} \cdot \left[ 10^{\alpha_f H_{max}} - 10^{\alpha_f H_B} \right] + 10^{\alpha_f H_B}) \\
                                                                                                            H' & = \frac{1}{\alpha_f} \log_{10}(\mathcal{U} \cdot \left[ 10^{\alpha_f H_{max}} - 10^{\alpha_f H_B} \right] + 10^{\alpha_f H_B}) \\
             \end{aligned}
    \end{aligned}
\end{equation}

One can now draw $N_{total}$ number of objects from the bright and faint distributions combined according to the fraction determined in Equation \ref{ap:eq7}. The differential, cumulative, and inverse transforms for the bright end faint end distributions are summarized in Table \ref{ap:tab1}.

\begin{deluxetable}{ccc} 
    \tablecaption{
        Summary of the differential, cumulative, and inverse transforms of the divot power law. \label{ap:tab1}
    }
    \tablehead{
        \colhead{ } & \colhead{$H < H_B$} & \colhead{$H \geq H_B$} 
    }
    \startdata
        Differential       &  $k_b \alpha_b \ln(10) 10^{\alpha_b H}$                            &  $k_f \alpha_f \ln(10) 10^{\alpha_f H}$                                                                                             \\
        Cumulative         &  $k_b 10^{\alpha_b H}$                                           &  $k_f \frac{1}{c} \frac{\alpha_b}{\alpha_f} 10^{\alpha_b H_B} \left( 10^{\alpha_f (H - H_B)} - 1 \right)$                           \\
        Inverse Transform  &  $\frac{1}{\alpha_b} \left( \log_{10}(\mathcal{U}) + \alpha_b H_B \right)$   &  $\frac{1}{\alpha_f} \log_{10}\left( \mathcal{U} \cdot \left( 10^{\alpha_f H_{max}} - 10^{\alpha_f H_B} \right) + 10^{\alpha_f H_B} \right)$  \\
    \enddata
    \tablecomments{
        $k_b = A \cdot 10^{-\alpha_b H_0}$, bright end normalization parameter; $k_f = B \cdot 10^{-\alpha_f H_B}$, faint end normalization parameter; $\alpha_b / \alpha_f$, bright/faint end slope parameter; $c$, slope parameter (ratio of number of objects in infinitesimally small bin pre-break to number of objects in infinitesimally small bin post-break); $H_B$, break absolute magnitude; $H_0$, normalization absolute magnitude; $H_{max}$, largest absolute magnitude to draw from; $\mathcal{U}$, uniform distribution $\in [0, 1]$
    }
\end{deluxetable}

\section{Sampling from a Multivariate Normal Distribution for Multiple Component Gaussian Mixture Models} \label{ap:2}

Sampling from a multivariate normal distribution with a given mean $\bm{\mu}$ and covariance $\bm{C}$ (i.e., $\bm{X} \sim \mathcal{N}(\bm{\mu}, \bm{C})$) is a straightforward implementation of the \texttt{numpy.random.multivariate\_normal} function. In the case of Gaussian Mixture Modeling, however, where there may be $K$ normal distributions within the overall target distribution to sample from, this function is not able to parse $>1$D $\bm{\mu}$ and $>2$D $\bm{C}$ arrays. In this instance, we instead first sample from a standard multivariate normal $\bm{Z} \sim \mathcal{N}(\bm{0},\bm{I})$, where $\bm{I}$ is the identity matrix. As the sample distribution $\bm{X}$ is simply an affine-transformed version of the sample distribution $\bm{Z}$, we can transform between the two by:

\begin{equation*}
    \bm{X} = \bm{A}\bm{Z} + \bm{\mu} 
\end{equation*}
where $\bm{A}$ is a transformation matrix that satisfies the relation $\bm{C} = \bm{A}\bm{A}^T$ (this arises from the fact that $\bm{C}$ = cov[$\bm{X}$] = $\mathbb{E}[(\bm{X} - \bm{\mu})(\bm{X}-\bm{\mu})^T] = \bm{A} \mathbb{E}[\bm{Z}\bm{Z}^T] \bm{A}^T = \bm{A}\bm{I}\bm{A}^T = \bm{A}\bm{A}^T$). The transformation matrix $\bm{A}$ can be found via Cholesky decomposition of the covariance matrix $\bm{C}$ (where $\bm{A}$ is a lower triangle matrix whose elements are all real and whose diagonal elements are positive), which is implemented via the \texttt{numpy.linalg.cholesky} function. This entire sampling method is wrapped in the GMM analysis code used in \cite{bernardinelli25}, and is publicly available for open use at \href{https://github.com/bernardinelli/gmm\_anyk}{https://github.com/bernardinelli/gmm\_anyk}.

\section{Yield Range for Absolute Magnitude Uncertainties} \label{ap:3}

\begin{figure}
    \centering
    \includegraphics[width=\linewidth]{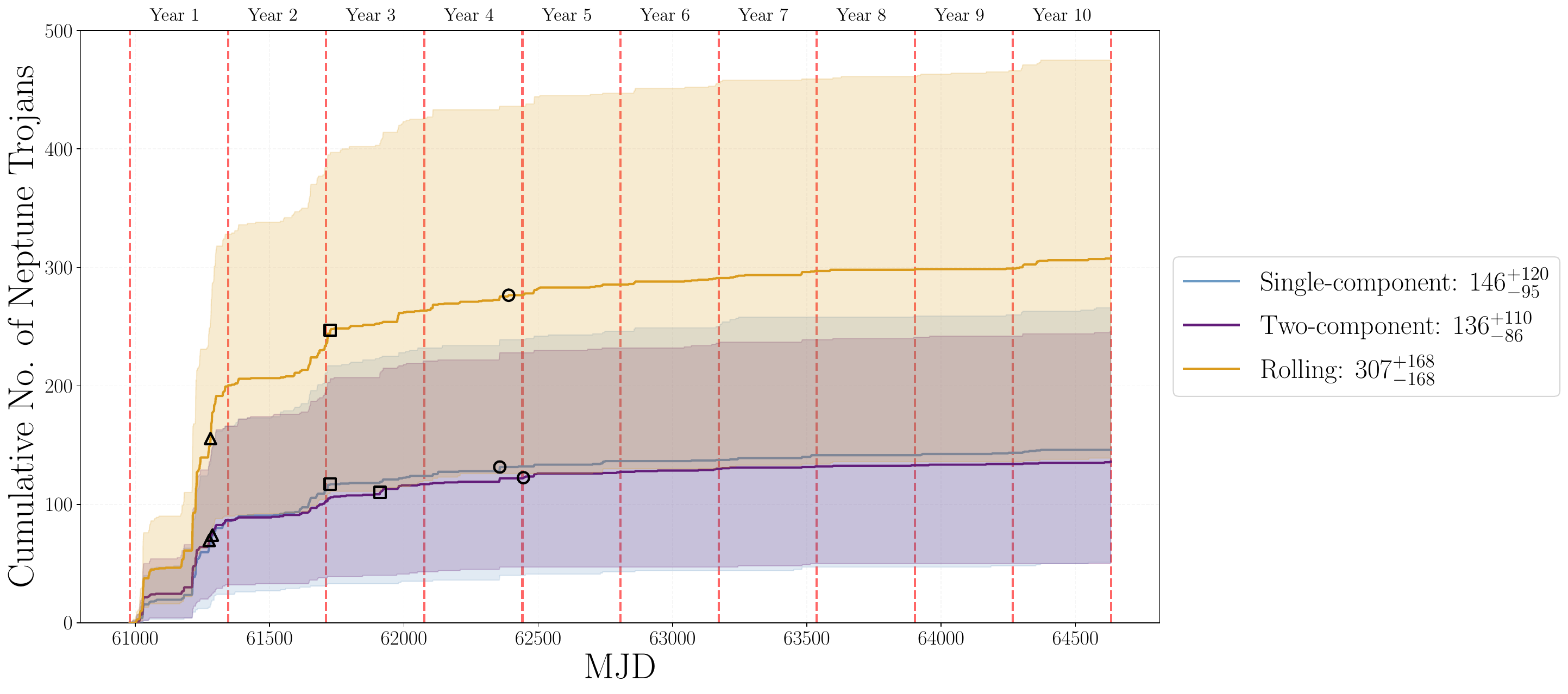}
    \caption{Cumulative histogram (with a bin size of one day) of the discovery rates over the 10 year LSST lifetime for the single-component (blue), two-component (purple), and rolling power law (orange) models. The points of 50\% (open triangle) 80\% (open square), and 90\% (open circle) completion are shown for each model. Red dashed vertical lines represent survey year start/end points. The ranges here represent the upper and lower estimate for the scaling parameters used in each model, described in Section \ref{sec:2.3.1}.}
    \label{fig:unc_yield}
\end{figure}

The yields for each NT absolute magnitude models is shown in Figure \ref{fig:unc_yield}, with the range on each representing the upper and lower estimate for the scaling parameters used in constructing each model (see Section \ref{sec:2.3.1}). Table \ref{tab:unc_yield} breaks down the yield per year for both the upper and lower estimates for years 1, 2, 5, and 10 of the LSST operations.

\begin{deluxetable}{c c c c c c}[htb]
\tablecaption{Number of Neptune Trojan discoveries by survey year for the lower and upper limit from each absolute magnitude model's scaling parameter described in Section \ref{sec:2.3.1}. As a reference, the Minor Planet Center only notes 31 Neptune Trojans.\label{tab:unc_yield}}
\tablewidth{\columnwidth}
\tablehead{
 & & \multicolumn{4}{c}{LSST Discovery Numbers}  \\ \cline{3-6}
 & & \colhead{1 yr} & \colhead{2 yrs} & \colhead{5 yrs} & \colhead{10 yrs}
}
\startdata
\multirow{3}{*}{Upper estimate} & Single-component & 166  & 211  & 246  & 266 \\
                                & Two-component    & 166  & 194  & 232  & 246 \\   
                                & Rolling          & 328  & 384  & 447  & 475 \\  \cline{1-6}   
\multirow{3}{*}{Lower estimate} & Single-component & 24   & 32   & 43   & 51  \\
                                & Two-component    & 32   & 38   & 47   & 50  \\   
                                & Rolling          & 90   & 105  & 130  & 139 \\   
\enddata
\end{deluxetable}





\bibliography{zrefs}{}
\bibliographystyle{aasjournalv7}



\end{CJK*}

\end{document}